\documentclass[pdflatex,sn-mathphys-num]{sn-jnl} 

\usepackage[utf8]{inputenc}
\usepackage[T1]{fontenc}
\usepackage[english]{babel}
\usepackage{amsmath,amssymb,amsfonts,color,graphicx,bm,times}

\begin{document}

\title[Article Title]{Solitary wave solutions, periodic and superposition solutions to the system of first-order (2+1)-dimensional Boussinesq's equations derived from the Euler equations for an ideal fluid model}

\author*[1]{\fnm{Piotr} \sur{Rozmej}}\email{piotr.rozmej@ifmpan.poznan.pl}

\author[2]{\fnm{Anna} \sur{Karczewska}}\email{a.karczewska@im.uz.zgora.pl}
\equalcont{These authors contributed equally to this work.}

\affil*[1]{\orgdiv{Institute of Molecular Physics}, \orgname{Polish Academy of Sciences}, \orgaddress{\street{M.~Smoluchowskiego 17}, \city{Pozna\'n}, \postcode{60-179},  \country{Poland}}}

\affil[2]{\orgdiv{Institute of Mathematics}, \orgname{University of Zielona G\'ora}, \orgaddress{\street{zafrana 4a}, \city{ielona G\'ora}, \postcode{65-516},  \country{Poland}}}

\abstract{This article concludes the study of (2+1)-dimensional nonlinear wave equations that can be derived in a model of an ideal fluid with  irrotational motion. In the considered case of identical scaling of the $x,y$ variables, obtaining a (2+1)-dimensional wave equation analogous to the KdV equation is impossible. Instead, from a system of two first-order Boussinesq equations, a non-linear wave equation for the auxiliary function $f(x,y,z)$ defining the velocity potential can be obtained, and only from its solutions can the surface wave form $\eta(x,y,t)$ be obtained. We demonstrate the existence of families of (2+1)-dimensional traveling wave solutions, including solitary and periodic solutions, of both cnoidal and superposition types.}



\maketitle

\section{Introduction} \label{intro}

Studies of nonlinear wave equations and their solutions have experienced an impressive boom in recent years as they find important applications in many areas of physics and technology. Most of the latest research focuses on wave phenomena in two-dimensional, three-dimensional, and even n-dimensional media.  
Therefore, functions describing wave propagation are written as (2+1)- or (3+1)-dimensional, with two or three spatial variables and one temporal variable.  
In most of these studies, mathematically constructed equations with arbitrary coefficients are considered, yielding exciting solutions, 
see e.g. 
\cite{CM1994,Waz2009,Zhai16,Zhang17,MZMa2018,Lou20,ZH2020,GHLM2020,Malik22,MaWL2021,SK-SM2024}.
Most of these equations are integrable. The authors present many interesting analytical solutions: solitons, multi-solitons, breathers, lumps, etc., sometimes using complex variables. The (2+1)-dimensional and (3+1)-dimensional equations used in these kinds of studies were not derived from fundamental laws of hydrodynamics but rather constructed by analogy to one-dimensional KdV-type equations or KP equation. 
These constructions often utilized integrability, symmetries, and conservation laws, see e.g. \cite{Waz2010,Khal2013,Adem16,Wang2019,Wang21,Tiwari22,FCH22,IN2022}. 
In his famous book \cite{Osborne}, Osborne mentions that the  KP, (2+1)-dimensional KP-Gardner, and (2+1)-dimensional Gardner  equations in their simple form are integrable by inverse scattering transform method (IST). 
The authors of these studies often claim that their results can be applied to explain wave behaviour in shallow waters. However, there is no evidence that equations of this type are good approximations to the general equations of hydrodynamics, and usually, no debate about their applicability. 

In \cite{Lannes}, Lannes reviews several models that have been derived for modeling shallow water flows, focusing on those of interest for applications to coastal oceanography. The derivations are based on the full hydrodynamic equations (the Euler equations for an ideal, incompressible, and inviscid fluid with a free surface) using depth-averaging and asymptotic expansions. The process typically involves assuming the fluid layer is shallow (horizontal length scales are large compared to the vertical), allowing expansion in small non-dimensional parameters representing the shallowness and weak nonlinearity.  
Within this methodology Horikis et.~al.~\cite{MS2021,S2022} derived (2+1)-dimensional extended KdV equation, the extended Kadomtsev-Petviashvili (eKP) equation, and the extended cylindrical KdV equation. 

Based on our experience with the (1+1)-dimensional nonlinear wave equations \cite{KRI,RKI,IKRR,RK,KRIR,KRBook,KRcnsns20}, which can be derived as approximations in the perturbation approach from the Euler equations for the irrotational flow of an ideal fluid, in 2022 we started work on the generalisation of this theory to the (2+1)-dimensional equations. 
In both our work on (1+1)-dimensional and (2+1)-dimensional equations, we have followed the method of ordering of small parameters used by Burde and Sergyeiev \cite{BS2013}.
In \cite{KRcnsns23,RKappl24}, we derived (2+1)-dimensional extensions of the KdV, fifth-order KdV, and Gardner equations using non-uniform scaling. Furthermore, we have shown that the Kadomtsev-Petviashvili (KP) equation follows directly from the (2+1)-dimensional KdV equation derived in the ideal fluid model. We have also shown that there are traveling wave-type solutions for these equations, analogous to the solutions of the corresponding (1+1)-dimensional equations. In \cite{RK2023NoDy,RK25WaveM}, we derived the (2+1)-dimensional extended KdV equation and the extended KP equation and their soliton, cnoidal, and superposition solutions of traveling wave-type.
In \cite{KR-BoussND}, we derived two first-order (2+1)-dimensional Boussinesq equations for the case of uniform scaling of horizontal coordinates. We have outlined a scheme for dealing with general boundary conditions, but have been unable to go further. In the present work, we have found several families of solutions to these equations in the form of traveling waves, analogous to the solutions of the classical KdV equation, i.e. soliton solutions and periodic cnoidal and superposition solutions.

The paper is organized as follows. Section \ref{model} introduces a system of Euler equations for an inviscid, incompressible fluid whose motion is irrotational. 
In section \ref{sec3}, we briefly recall results obtained for (2+1)-dimensional equations derived for cases of non-uniform scaling \cite{KRcnsns23,RKappl24,RK2023NoDy,RK25WaveM}. The main part of the paper, containing the new equations and their traveling wave solutions, is section \ref{Bouss-abg1}. The article closes with conclusions.

\section{Model description} \label{model}
Consider an inviscid and incompressible fluid whose motion is irrotational in a huge container with a flat, impenetrable bottom. 
In dimensional variables, the set of hydrodynamical equations consists of the Laplace equation for the velocity potential $\phi(x,y,z,t)$ and boundary conditions at the free surface and the bottom
\begin{eqnarray}  \label{g1}
\phi_{xx} + \phi_{yy} + \phi_{zz}&=&0, \quad \mbox{in~the~volume}, \\ \label{g2}
\phi_z - ( \eta_x \phi_x +  \eta_y \phi_y + \eta_t) &=& 0, \quad \mbox{at} \quad z=H+A\,\eta,  \\ \label{g3}
\phi_t + \frac{1}{2}(\phi_x^2+\phi_y^2+\phi_z^2) +g \eta &=& 0,  \quad \mbox{at}  \quad z=H+A\,\eta, \\ \label{g4}
\phi_z  &=& 0,  \quad \mbox{at} \quad z=0.
\end{eqnarray}
Here, $ \eta(x,y,t)$ denotes the surface profile function, $g$ is the gravitational acceleration, $A$ is the amplitude of surface distortions from the equilibrium shape (flat surface), and $H$ is the fluid depth.
Indexes denote partial derivatives, i.e.\ $\phi_{x}\equiv \frac{\partial \phi}{\partial x}, ~\eta_{y}\equiv \frac{\partial \eta}{\partial y}, ~\phi_{xx}\equiv \frac{\partial^{2} \phi}{\partial x^{2}}$, and so on. 
Equations (\ref{g2})-(\ref{g3}) are kinematic and dynamic boundary conditions at the unknown surface, respectively.

The next step consists of introducing a standard scaling to dimensionless variables (in general, it could be different in $x$, $y$, and $z$ directions)
\begin{align} \label{bezw}
\tilde{x} & =x/L_{x}, \quad \tilde{y}= y/L_{y},\quad \tilde{z}= z/H, \quad \tilde{t}= t/(L_{x}/\sqrt{gH}), \quad 
\tilde{ \eta}  =  \eta/A,\quad \tilde{\phi}= \phi /(L_{x}\frac{A}{H}\sqrt{gH}).\end{align}
Here, $L_{x}$ and $L_{y}$ are the scaling factors in the $x$ and $y$ directions,  respectively. For (1+1)-dimensio\-nal equations, $L_{x}$ is often understood as the average wavelength. The theory is intended to apply to long waves in shallow water, so the horizontal scaling factors $L_{x},L_{y}$ should be distinctly larger than the water depth $H$, which is the scaling factor in the vertical direction. In general,  $L_{y}$ should be in the same order as $L_{y}$, but not necessarily equal. 
Then the set (\ref{g1})-(\ref{g4}) takes in scaled variables the following form (here and next, we omit the tilde signs)
\begin{align}  \label{G1}
\beta \phi_{xx} + \gamma\phi_{yy} + \phi_{zz}&= 0,  \quad \mbox{in~the~volume} \\ \label{G2}
  \eta_t +\alpha ( \eta_x\phi_x+\frac{\gamma}{\beta} \eta_y\phi_y)-\frac{1}{\beta}\phi_z &= 0,  
 \quad \mbox{for~~} z  =1+\alpha\, \eta, \\ \label{G3}
\phi_t + \frac{1}{2}\alpha \left(\phi_x^2+\frac{\gamma}{\beta}\phi_y^2+\frac{1}{\beta}\phi_z^2\right) +  \eta &= 0,   
 \quad \mbox{for~~} z =1+\alpha\, \eta,  \\ \label{G4}
\phi_z   &= 0, 
 \quad \mbox{for~~} z =0. 
\end{align}
Besides standard small parameters $\alpha=\frac{A}{H}$,  $\beta=\left(\frac{H}{L_{x}}\right)^2$, we introduced another one defined as $\gamma=\left(\frac{H}{L_{y}}\right)^2$.
In the perturbation approach, all these parameters, $\alpha,\beta,\gamma$, are assumed to be small but not necessarily of the same order.  
The standard perturbation approach to the system of Euler's equations (\ref{G1})-(\ref{G4}) consists of the following steps. First, the velocity potential is sought in the form of power series in the vertical coordinate
\begin{equation} \label{Szer}
\phi(x,y,z,t)=\sum_{m=0}^\infty z^m\, \phi^{(m)} (x,y,t),
\end{equation}
where ~$\phi^{(m)} (x,y,t)$ are yet unknown functions. The Laplace equation (\ref{G1}) with the boundary condition at the bottom (\ref{G4}) determines $\phi$ in the form which involves only one unknown function with the lowest $m$-index, $f(x,y,t):=\phi^{(0)} (x,y,t)$ and its space derivatives. Therefore,
\begin{align} \label{Szer1}
 \phi(x,y,z,t) & =\sum_{m=0}^\infty \frac{(-1)^m}{(2m)!} z^{2m}\, (\beta\partial_{xx}+\gamma\partial_{yy})^m  f(x,y,t)  . \end{align}
The explicit form of this velocity potential reads as
\begin{align} \label{pot8}
\phi = f & -\frac{1}{2} z^2 (\beta f_{2x}+\gamma f_{2y}) \nonumber   
+ \frac{1}{24} z^4 (\beta^2 f_{4x}+2\beta\gamma f_{2x2y}+\gamma^2f_{4y}) \\ &
 - \frac{1}{720}\beta^3 z^6 (f_{6x}+3f_{4x2y}+3f_{2x4y}+f_{6y})
+ \cdots  
\end{align}

Next, the velocity potential is substituted into kinematic and dynamic boundary conditions at the unknown surface (\ref{G2})-(\ref{G3}). Retaining only terms up to a given order, one obtains the Boussinesq system of two equations for unknown functions $ \eta,f$ valid only up to a given order in small parameters. In principle, for a flat bottom, the Boussinesq equations may be obtained up to arbitrary order. The resulting equations, however, depend substantially on the ordering of small parameters. If the bottom is not flat, the Boussinesq equations can be obtained up to second order at most \cite{KRcnsns20}.

In 2013, Burde and Sergyeyev \cite{BS2013} demonstrated that for the case of (1+1)-dimensional and the flat bottom, the KdV, the extended KdV, fifth-order KdV, and Gardner equations can be derived from the same set of Euler's equations (\ref{G1})-(\ref{G4}). Different final equations result from the different ordering of small parameters and consistent perturbation approach up to first or second order in small parameters.

In the next section, we recall results of derivations of (2+1)-dimensional extensions of the KdV, fifth-order KdV, Gardner (KdV-mKdV), and the extended KdV equations given in \cite{KRcnsns23,RKappl24,RK2023NoDy,RK25WaveM,KR-BoussND}. These articles also contain travelling wave solutions to these equations.

\section{Non-uniform scaling, $\gamma=0(\beta^{2})$ or $\gamma=0(\alpha^{2})$, $\alpha=0(\beta)$. (2+1)-dimensional nonlocal KdV, fifth-order KdV, Gardner (KdV-mKdV), extended KdV equations and their travelling wave solutions} \label{sec3}

In the series of papers \cite{KRcnsns23,RKappl24,RK2023NoDy,RK25WaveM,KR-BoussND} we have undertaken the task of generalising the KdV, fifth-order KdV, Gardner (KdV-mKdV), and the extended KdV equations to (2+1)-dimensions. 
In the following, we will recall how these equations were derived and their solutions in terms of traveling waves. 

\subsection{(2+1)-dimensional nonlocal KdV equation and its travelling solutions}  \label{sec3.1}

Begin with the case when $~\alpha\approx\beta, ~\gamma\approx\beta^2$.
Inserting the velocity potential (\ref{pot8}) into the kinematic boundary condition at the surface (\ref{G2}) and neglecting terms higher than the first order in small parameters yields
\begin{equation} \label{g2G2}
\eta_{t}+f_{xx} + \alpha (\eta f_{x})_{x} -\frac{1}{6} \beta f_{4x} +\frac{\gamma}{\beta} f_{yy}=0.
\end{equation}
Analogous steps with the dynamic  boundary condition at the surface (\ref{G3}) lead to the first order equation 
\begin{equation} \label{g2G3}
\eta+f_{t} + \frac{1}{2}\alpha f_{x}^{2} -\frac{1}{2} \beta f_{xxt}  =0.
\end{equation}
Equations (\ref{g2G2})-(\ref{g2G3}) constitute the first-order Boussinesq's equations for the case when $\alpha \approx \beta$, $\gamma \approx \beta^{2}$ . Despite the assumption that $\gamma$ is of the second order, the term $\frac{\gamma}{\beta} f_{yy} $ appears in the Boussinesq equation (\ref{g2G2}) as the first order one. 

Next, we applied a standard method for making the Boussinesq equations (\ref{g2G2})-(\ref{g2G3}) compatible, which in (1+1)-dimensions leads to the Korteweg-de Vries equation.
By differentiating over $x$ the equation (\ref{g2G3}) and denoting $f_{x}=w, \hspace{1ex} f=\partial_{x}^{-1}(w), \hspace{1ex} f_{yy}=\partial_{x}^{-1} (w_{yy}) $ we can write the equations (\ref{g2G2})-(\ref{g2G3}) in the form
\begin{align} \label{et}
\eta_{t}+w_{x} & + \alpha (\eta w)_{x} -\frac{1}{6} \beta w_{3x} +\frac{\gamma}{\beta}\,\partial_{x}^{-1} (w_{yy}) = 0,\\ \label{wt}
w_{t}+\eta_{x} & + \alpha w w_{x} -\frac{1}{2} \beta w_{xxt} =0.
\end{align}
Here and hereafter, the $\partial_{x}^{-1}$ operator is defined as
\begin{equation} \label{part1}
\partial_{x}^{-1}( w )=\int_{-\infty}^{x} w(x',y,t)\, dx'.
\end{equation} 
Equation (\ref{et}) has a nonlocal form. When the problem is reduced to (1+1)-dimensions ($u,w$ not dependent on $y$)  
equations (\ref{et})-(\ref{wt}) reduce to the classical Boussinesq equations, leading to the KdV equation. Note, that the new term $\frac{\gamma}{\beta} \partial_{x}^{-1} (w_{yy})$ is a first-order one because $\frac{\gamma}{\beta}\approx \beta$. Then in zeroth order, the following holds
\begin{equation} \label{0g2}
\eta_{t}+w_{x}=0, \quad w_{t}+\eta_{x}=0, \qquad \mbox{implying} 
\qquad w=\eta, \quad w_{t}=-w_{x}, \quad \eta_{t}=-\eta_{x}.
\end{equation} It is worth emphasizing that zeroth-order relations (\ref{0g2}) are the same as in the one-dimensional case. These relations, allows us to replace some $t$-derivatives by $-x$-derivatives and are crucial in deriving both (1+1)-dimensional and (2+1)-dimensional KdV-type equations.   

To make equations (\ref{et})-(\ref{wt}) compatible we postulate $w$ in the following form
\begin{equation} \label{1w}
w = \eta +\alpha Q^{(a)}+\beta Q^{(b)}+\frac{\gamma}{\beta} Q^{(g)} ,
\end{equation}
where $\alpha Q^{(a)}, \beta Q^{(a)}, \frac{\gamma}{\beta} Q^{(g)}$ are first-order corrections. Inserting (\ref{1w}) into  (\ref{et})-(\ref{wt}), replacing $t$-derivatives by $-x$-derivatives (according to (\ref{0g2})), one can obtain differential equations for the correction functions  $Q^{(a)}, Q^{(b)}, Q^{(g)}$.
Solving these equations one obtains the proper form of $w$ (\ref{1w}), which after substituting to Boussinesq's equtions (\ref{et})-(\ref{wt}), and leaving out terms up to the first order, reduces each to the same {\bf (2+1)-dimensional non-local KdV equation} in the fixed frame \cite[Eq.~(32)]{KRcnsns23}
\begin{equation} \label{et11}
 \eta_{t} +\eta_{x} +\frac{3}{2} \alpha \eta \eta_{x}+ \frac{1}{6} \beta  \eta_{xxx} + \frac{1}{2} \frac{\gamma}{\beta} \int \eta_{yy}\, dx =0.
 \end{equation}
When $y$-derivatives are zero (or $\gamma=0$), equation (\ref{et11}) reduces to the usual KdV equation in a fixed frame.
Differentiating (\ref{et11}) over $x$ yields
\begin{equation} \label{kpt}
\frac{\partial }{\partial x}\left(  \frac{\partial \eta}{\partial t} + \frac{\partial \eta}{\partial x} +\frac{3}{2} \alpha\, \eta \frac{\partial \eta}{\partial x} +\frac{1}{6} \beta\, \frac{\partial^{3} \eta}{\partial x^{3}}  \right)= -\lambda\, \frac{\partial^{2} \eta}{\partial y^{2}}, \quad \mbox{where} \quad \lambda=\frac{1}{2} \frac{\gamma}{\beta}. 
 \end{equation}
Equation (\ref{kpt}) represents a general form of the Kadomtsev-Petviashvili equation in a fixed reference frame. When $\alpha=\beta$, the 
transformation ~$ \hat{x} =\sqrt{\frac{3}{2}}(x-t), \quad \hat{t} =\frac{1}{4} \sqrt{\frac{3}{2}}\,\alpha t, ~\mbox{and} ~\hat{y}= y$~ reduces (\ref{kpt})  to the classical KP equation \cite{KP}.

In \cite{KRcnsns23}, we showed that the equation (\ref{et11}) has families of travelling wave solutions analogous to those of the one-dimensional KdV equation, namely:
\begin{itemize}
\item Solitary waves in the form ~$\eta = A\,\text{sech}^{2}(k x\pm l y\pm\omega t)$.
\item Periodic cnoidal waves in the form ~$\eta = A\,\text{cn}^{2}((k x\pm l y\pm \omega t),m)+ C$.
\item Periodic superosition waves in the form\\ $\eta =\frac{A}{2}\left[\text{dn}^{2}((k x\pm l y\pm \omega t),m)\pm\sqrt{m}\,\text{cn}((k x\pm l y\pm \omega t),m)\,\text{dn}((k x\pm l y\pm \omega t),m) \right]+ C$.
\end{itemize}
The signs $\pm$ in the argument are irrelevant, they only determine the direction of wave propagation. 

\subsection{(2+1)-dimensional nonlocal fifth-order KdV equation and its travelling solutions}  \label{sec3.2}

Consider the case when $~\alpha\approx\beta^{2}, ~\gamma\approx\beta^2$.
When the theory is applied to thin fluid layers, surface tension can play an important role. Expressions related to surface tension appear in the dynamic boundary condition (\ref{G3}), which then takes the following form \cite[Eq.~(75)]{KRcnsns23}
$$
\qquad \phi_t + \frac{1}{2}\alpha \left(\phi_x^2+\frac{\gamma}{\beta}\phi_y^2+\frac{1}{\beta}\phi_z^2\right) + \eta -\tau\,(\beta\eta_{xx}+\gamma \eta_{yy})
= 0,\quad \mbox{for} \quad z=1+\alpha\eta . \quad\quad (8a) $$
Here,  $\tau=\frac{T}{\varrho g H^{2}}$ is the Bond number ($T$ is the surface tension coefficient, $\varrho$ is the density of the fluid, and $g$ is the gravitational acceleration).  For ordinary shallow water waves (depths on the order of meters, $\tau \in (10^{-8}-10^{-6})$), surface tension effects can be safely neglected by setting $\tau=0$. However, when $H$ is of the order of millimetres $\tau$ can reach values close to 1. Then, the terms originating from surface tension cannot be neglected.
Substituting the velocity potential (\ref{pot8}) into (\ref{G2}) and (8a), and retaining terms up to the second order in $\beta$ we obtained the following set of the Boussinesq equations \cite[Eqs.~(77)-(78)]{KRcnsns23}
\begin{align} \label{bu1-6} \eta_{t} +w_{x} &- \frac{1}{6}\beta\,w_{3x} +\frac{1}{120} \beta\, w_{5x}+\frac{\gamma}{\beta}\partial_{x}^{-1} (w_{yy})  + \alpha \left(\eta w\right)_{xx} -\frac{1}{3}\gamma\,w_{x2y} =0 ,\\  \label{bu2-6} 
w_{t}+\eta_{x} &- \beta\left( \frac{1}{2}\,w_{2xt}+\tau \eta_{3x} \right) + \alpha\, w w_{x}+\frac{1}{24}\beta^{2}\,w_{4xt}-\gamma\left(\frac{1}{2}w_{2yt}+\tau \eta_{x2y}\right)  =0. \end{align}  
In (\ref{bu1-6})-(\ref{bu2-6}), 
zeroth-order terms have the same form as in (\ref{et})-(\ref{wt}), so relations (\ref{0g2}) hold. Therefore, one can use the same procedure to eliminate $w$ and make equations (\ref{bu1-6})-(\ref{bu2-6}) compatible. 
The final resul is
\begin{align} \label{gkdv5}
\eta_{t} +\eta_{x} & +\beta \left(\frac{1-3\tau}{6}\right)\eta_{3x}+\frac{\gamma}{2\beta} \partial_{x}^{-1} (\eta_{yy}) + \frac{3}{2}\alpha\, \eta \eta_{x}  
+\beta^{2} \left(\frac{19-30\tau-45\tau^{2}}{360}\right)\eta_{5x} \\ &
+\gamma\left(\frac{1-3\tau}{4} \right) \eta_{x2y}  -
\frac{1}{8} \frac{\gamma^{2}}{\beta^{2}} \partial_{x}^{-3} (\eta_{4y}) =0 .  \nonumber 
\end{align} 
Equation (\ref{gkdv5}) is 
{\bf (2+1)-dimensional fifth-order Korteweg-de Vries equation }. When $y$-derivatives are zero (or $\gamma=0$), equation (\ref{gkdv5}) reduces to the usual fifth-order KdV equation.

In \cite{RKappl24}, we showed that there exist soliton solutions to equation (\ref{gkdv5}) in the form
\begin{equation} \label{ssol5}
\eta(x,y,t)=A\, \text{sech}^{4}(k x+l y-\omega t),\end{equation}
analogous to soliton solutions to one-dimensional fifth-order Korteweg-de Vries equation.

\subsection{(2+1)-dimensional nonlocal Gardner  equation and its travelling solutions}  \label{sec3.3}

Consider the case when $~\beta\approx\alpha^{2}, ~\gamma\approx\alpha^2$.
The Boussinesq equations obtained by retaining terms up to the second order take the following form (surface tension effects included) \cite[Eqs.~(94)-(95)]{KRcnsns23}
\begin{align} \label{BGa13}
\eta_{t} + w_{x} & +\frac{\gamma}{\beta}\partial_{x}^{-1} (w_{yy}) +\alpha (\eta w)_{x} + \frac{\alpha\gamma}{\beta}\left(\eta_{y}\, \partial_{x}^{-1} (w_{y}) +\eta\,\partial_{x}^{-1} (w_{yy}) \right)-\frac{1}{6} \beta w_{3x} =0,\\  \label{BGa23}
w_{t} + \eta_{x} & + \alpha w w_{x} +\frac{\alpha\gamma}{\beta}\, w_{y}\,\partial_{x}^{-1} (w_{y}) - \frac{1}{2}\beta\left( w_{xxt} +2\tau \eta_{xxx}\right) =0.
\end{align}
The form of zero-order terms makes it possible to use relation (\ref{0g2}) and, like in previous cases, reduce equations (\ref{BGa13})-(\ref{BGa23}) to the same nonlinear wave equation
\begin{align} \label{BGarEq}
\eta_{t}& + \eta_{x}  +\frac{3}{2}\alpha \eta\eta_{x} +\frac{\gamma}{2\beta}\partial_{x}^{-1} (\eta_{yy}) - \frac{3}{8}\alpha^{2} \eta^{2}\eta_{x}
+\frac{1}{6} \beta (1-3\tau) \eta_{3x}  \\ &+
 \frac{\alpha\gamma}{\beta}\left(\frac{1}{8} \partial_{x}^{-1} \left(\eta_{y}^2 +\eta \eta_{yy}\right)+\frac{1}{8}\eta\, \partial_{x}^{-1} (\eta_{yy})  +\eta_{y}\partial_{x}^{-1} (\eta_{y}) -\frac{1}{2} \eta_{x} \partial_{x}^{-2} (\eta_{yy}) \right)  
 - \frac{1}{8} \frac{\gamma^{2}}{\beta^{2}} \partial_{x}^{-3} (\eta_{4y}) =0 .   \nonumber
\end{align}
Equation (\ref{BGarEq}) constitutes the  {\bf (2+1)-dimensional Gardner equation} (sometimes referred to as the KdV-mKdV equation).

In \cite{RKappl24}, we discuss in details the soliton solutions to equation (\ref{gkdv5}) in the form
\begin{equation} \label{ssol6}
\eta(x,y,t)= \frac{A_{1}}{1\pm A_{2} \cosh(k x+l y-\omega t)},\end{equation}
analogous to soliton solutions to the one-dimensional Gardner equation.

\subsection{(2+1)-dimensional extended KdV equation and its travelling solutions}  \label{sec3.4}

Consider the case when $~\alpha\approx\beta, ~\gamma\approx\beta^2$, like in Sec.~\ref{sec3.1}, but extend the calculations to second order in small parameters. On the way to the final equations, some results obtained for the first-order equations can be used.

The Boussinesq equations obtained from (\ref{G2})-(\ref{G3}) by substituting  the velocity potential (\ref{pot8}) and retaining terms up to the second order take the following form (surface tension is now neglected) \cite[Eqs.~(40)-(41)]{RK25WaveM}
\begin{align} \label{r7w}
\eta_{t} +w_{x} & + \alpha (\eta w)_{x} +\frac{\gamma}{\beta} \partial_{x}^{-1} (w_{yy})
-\frac{1}{6}\beta \, w_{3x}\\ &  -\frac{1}{2} \alpha\beta (\eta\, w_{2x})_{x} + \frac{\alpha\gamma}{\beta} \left(\eta\, \partial_{x}^{-1} (w_{y})\right)_{y} -\frac{1}{3} \gamma \,w_{xyy}+\frac{1}{120}\beta^2 w_{5x} =0, \nonumber 
\\ \label{r8w}
w_{t} +\eta_{x} & + \alpha\,w w_{x} -\frac{1}{2}\beta\, w_{xxt} \\ &
  +\alpha\beta \left(\!  - (\eta\, w_{xt})_{x} +\frac{1}{2}\,w_{x} w_{xx} -\frac{1}{2} \,w w_{3x}\! \right) + \frac{\alpha\gamma}{\beta} w_{y}\, \partial_{x}^{-1} (w_{y}) -\frac{1}{2} \gamma\, w_{yyt}+\frac{1}{24}\beta^2 w_{4xt} =0. \nonumber
\end{align}
In this case, extending search for second-order corrections one can finally obtain from both (\ref{r7w}) and (\ref{r8w}) equations the same {\bf (2+1)-dimensional extended Korteweg-de Vries equation} valid up to second-order terms
\begin{align} \label{2dKdV2}
\eta_{t} & +\eta_{x} + \frac{3}{2} \alpha \eta \eta_{x} +\frac{1}{6}\beta \eta_{3x}
+\frac{1}{2} \frac{\gamma}{\beta} \partial_{x}^{-1} (\eta_{yy}) \\ & 
-\frac{3}{8} \alpha^2 \eta^2 \eta_{x} +\alpha\beta\left( \frac{23}{24}\, \eta_{x} \eta_{xx}+\frac{5}{12}\, \eta\eta_{3x}\right)+ \frac{19}{360}\beta^2 \eta_{5x} +\frac{1}{4}\gamma\,\eta_{xyy}
\nonumber \\ & +
\frac{\alpha\gamma}{\beta}\left(\frac{3}{4} \partial_{x}^{-1} \left(\eta_{y}^2 + \eta \eta_{yy}\right)  +\frac{5}{4}\, \eta\, \partial_{x}^{-1} (\eta_{yy}) 
   -\frac{1}{2}\, \eta_{x} \, \partial_{x}^{-2} (\eta_{yy}) \right) 
   -\frac{1}{8}\frac{\gamma^{2}}{\beta^{2}}\, \partial_{x}^{-3} (\eta_{4y})  =0. \nonumber
\end{align}
It is worth noting that for single space dimension ($\frac{\partial \eta}{\partial y}=0$) the equation (\ref{2dKdV2}) reduces to the well-known extended Korteweg-de Vries equation (eKdV) \cite{ RKI, IKRR, RK, KRIR, KRBook, BS2013,MS90}
\begin{align} \label{KdV2}
\eta_{t} & +\eta_{x} +\frac{3}{2} \alpha \eta \eta_{x} +\frac{1}{6}\beta \eta_{3x} -\frac{3}{8} \alpha^2 \eta^2 \eta_{x} +\alpha\beta\left( \frac{23}{24}\, \eta_{x} \eta_{xx}+\frac{5}{12}\, \eta\eta_{3x}\right)+ \frac{19}{360}\beta^2 \eta_{5x}= 0. 
\end{align}

By differentiating the (2+1)-dimensional extended Korteweg-de Vries equation (\ref{2dKdV2}) with respect to $x$ we obtain the extended KP (eKP) equation 
\begin{align} \label{eKP1}
& \left(\eta_{t}+\eta_{x}\right)_{x} + \left(\frac{3}{2} \alpha \eta \eta_{x} +\frac{1}{6}\beta \eta_{3x}\right)_{x} +\frac{1}{2} \frac{\gamma}{\beta} \eta_{yy} \\ & 
+\left( -\frac{3}{8} \alpha^2 \eta^2 \eta_{x} +\alpha\beta\, ( \frac{23}{24}\, \eta_{x} \eta_{xx}+\frac{5}{12}\, \eta\eta_{xxx}) + \frac{19}{360}\beta^2 \eta_{5x}\right)_{x} +\frac{1}{4}\gamma\,\eta_{xxyy}
\nonumber \\ & +
\frac{\alpha\gamma}{\beta} \left( \frac{3}{4}\eta_{y}^2 + 2 \eta \eta_{yy}
  +\frac{3}{4}\, \eta _{x}\,\partial_{x}^{-1} (\eta_{yy})  
   -\frac{1}{2}\, \eta_{xx}\, \partial_{x}^{-2} (\eta_{yy})  \right)
 -\frac{1}{8}\frac{\gamma^{2}}{\beta^{2}}\, \partial_{x}^{-2} (\eta_{yyyy}) =0. \nonumber
\end{align}
It is worth noting that, within the same model of an ideal fluid in an irrotational motion, the authors of papers \cite{MS2021,S2022} have recently obtained (among other important results) an eKP equation almost identical to ours. Minor differences are due to different definitions of small parameters in the perturbation approach.

In \cite{RK2023NoDy}, we showed that the (2+1)-dimensional extended Korteweg-de Vries equation (\ref{2dKdV2}) possesses the same families of travelling solutions 
as one dimensional extended KdV equation (\ref{KdV2}) and (2+1)-dimensional KdV equation (\ref{et11}), namely soliton solutions, periodic cnoidal solutions and periodic superposition solutions. These solutions are simultaneously the solutions to the eKP equation (\ref{eKP1}).

\section{Uniform scaling of horizontal variables. 
(2+1)-dimensional first order Boussinesq's equations for ~$\alpha\approx\beta=\gamma$} \label{Bouss-abg1}

(2+1)-dimensional first order Boussinesq's equations for $\alpha\approx\beta \approx\gamma\ll 1$ are, see \cite[Eqs.~(18)-(19)]{KR-BoussND}
\begin{align} \label{BR7}
\eta_t +  f_{xx} \textcolor{blue}{\,+ \frac{\gamma}{\beta}f_{yy} } &
 +\alpha \left(\left(\eta f_x\right)_x+  \frac{\gamma}{\beta}\left(\eta f_y\right)_y \right) 
-\frac{1}{6}\,\beta \left(f_{4x}+2 \frac{\gamma}{\beta} f_{2x2y} +\frac{\gamma^{2}}{\beta^{2}}f_{4y}\right) = 0,   \\  \label{BR8} 
\eta  +f_{t}\hspace{8.5ex} &+\frac{1}{2} \alpha \left(f_{x}^{2} + \frac{\gamma}{\beta} f_{y}^{2}  \right)  -\frac{1}{2}\,\beta \left(f_{xxt} + \frac{\gamma}{\beta} f_{yyt}\right) =0.
\end{align}
For large water areas, it is most natural to scale the horizontal coordinates $x,y$ equally, which means that $\gamma =\beta$ and only two small parameters remain. Then the equations (\ref{BR7})-(\ref{BR8}) take the following form 
\begin{align} \label{BR7a}
\eta_t +  f_{xx} \textcolor{blue}{ \,+ f_{yy} } &
 +\alpha \left(\left(\eta f_x\right)_x+ \left(\eta f_y\right)_y \right) 
-\frac{1}{6}\,\beta \left(f_{4x}+2 f_{2x2y} +f_{4y}\right) = 0,   \\  \label{BR8a} 
\eta  +f_{t}\hspace{6.5ex} &+\frac{1}{2} \alpha \left(f_{x}^{2} +  f_{y}^{2}  \right)  -\frac{1}{2}\,\beta \left(f_{xxt} + f_{yyt}\right) =0.
\end{align}

In contrast to cases where $\gamma\approx \beta^{2}$ or $\gamma\approx \alpha^{2}$ considered in \cite{KRcnsns23,RKappl24,RK2023NoDy} it is not possible to make equatios (\ref{BR7})-(\ref{BR8}) compatible and consequently 
obtaining a single nonlinear wave equation for $\eta(x,y,t)$.  
In the present case, when  $\gamma$ is of the same order as $\alpha$ and $\beta$, the term $f_{yy}$ appears in zero-order equation (highlighted in blue in (\ref{BR7}) and (\ref{BR7a})), preventing relations (\ref{0g2}) and effectively making this impossible.

However, obtaining the first-order partial differential equation (PDE) for the function $f$ is possible. Inserting into (\ref{BR7a})
\begin{equation} \label{et1}
\eta   = -\left[f_{t} +\frac{1}{2} \alpha \left(f_{x}^{2} + f_{y}^{2}  \right)  -\frac{1}{2}\,\beta \left(f_{xxt} + f_{yyt}\right) \right]
\end{equation}
obtained from (\ref{BR8a}) and  retaining only terms up to first order yields
\begin{align} \label{f1ord}
 f_{xx}+ f_{yy}-f_{tt} &
 -\alpha \left[ f_{t}\left( f_{xx}+  f_{yy} \right)+\left(f_{x}^{~2} + f_{y}^{~2}\right)_{t}\right] \\ & \nonumber
 -\beta\left[\frac{1}{6}\left(f_{4x}+2 f_{2x2y} + f_{4y}\right)-\frac{1}{2}\left( f_{xxtt} + f_{yytt}\right)\right]=0.
\end{align}
If the solution $f(x, y,t)$ to (\ref{f1ord}) is known, equation (\ref{et1})
supplies the surface profile function $\eta(x, y,t)$. We have sketched out the steps for obtaining a general solution to the equation (\ref{f1ord}) in \cite{KR-BoussND} but have not gone any further because of the difficulty of the problem. 

Solutions of other cases of (2+1)-dimensional KdV-type equations found in \cite{KRcnsns23,RKappl24,RK2023NoDy} suggest searching for solutions in the form of travelling waves, that is
\begin{equation} \label{twe}
 f(x,y,t)=f(\xi), \qquad  \eta(x,y,t)=\eta(\xi), \qquad \mbox{where} \qquad \xi= k x \pm l y\pm \omega t.
\end{equation}
Since the signs in ~$\xi= k x \pm l y\pm \omega t$ affect only the direction of propagation of the wave and not its shape, we will continue using the formula ~$\xi= k x +l y- \omega t$ without loss of generality. The same argument allows us to limit our calculations to only nonnegative values of ~$k,l,\omega$~ parameters.

For the travelling wave (\ref{twe}), equation (\ref{f1ord}) becomes an ordinary differential equation (ODE) of the form 
\begin{equation} \label{rf1}
\left( q^2-\omega ^2\right)f''(\xi)
+3 \alpha  \omega q^2 f'(\xi ) f''(\xi) -\frac{1}{6}
\beta  q^{2} (q^2 -3\omega^2) f^{(4)}(\xi)=0
\end{equation}
or  
\begin{equation} \label{rf1a}
a f''+ 2 b f' f'' +c \, f^{(4)} =0 ,
\end{equation}
where
\begin{equation} \label{abc}
a=q^2 -\omega ^2, \quad  b=\frac{3}{2}\alpha \omega q^{2}, \quad c=-\frac{1}{6}
\beta  q^{2} (q^2 -3\omega^2), 
\quad \mbox{where} \quad q^{2}=k^2+ l^2.
\end{equation}

Integration of equation (\ref{rf1}) gives
\begin{equation} \label{rf1aa}
a f'+b (f')^{2}+ c\, f^{(3)} = r,
\end{equation}
where $r$ is an integration constant. 
Denote $F(\xi):=f'(\xi)$. Then equation (\ref{rf1aa}) reads as
\begin{equation} \label{rf1c}
a F+b F^{2}+ c\, F'' = r. \end{equation} 
Multiplication the equation (\ref{rf1c}) by $F'$ and intgration with respect to $\xi$ yields
\begin{equation} \label{rf1b}
(F')^{2} = -\frac{2b}{3c} F^{3} -\frac{a}{c} F^{2}+ \frac{r}{c} F +s .
\end{equation}

Equations of the type (\ref{rf1b}) appear when solving the KdV equation, see \cite{Whit,Ding} and when solving the relativistic Binet equation \cite{HZ2009,NR2009,NBRR2013}.

\subsection{Case 1, integration constants $r=s=0$. Solitary waves.}  \label{cas1}

Let us begin with the simplest case when both integration constants are set ~$r=s=0$.
This corresponds to solitary wave, when both $\eta(\xi),f(\xi)\to 0$ when $\xi\to \pm\infty$ (derivatives of $\eta(\xi),f(\xi)$ satisfy the same limits). Then 
\begin{equation} \label{rf1c1}
(F')^{2} = F^{2}\left(-\frac{2b}{3c} F -\frac{a}{c}\right) = F^{2}\left( \frac{6\alpha \omega}{\beta (q^{2}-3\omega^{2})} F +\frac{6(q^2 -\omega^2)}{\beta q^{2}(q^{2}-3\omega^{2})} \right).
\end{equation}

Equation (\ref{rf1c}) can be written as
\begin{equation} \label{rf1d}
\frac{\beta (q^{2}-3\omega^{2})}{6\alpha \omega}\,(F')^{2}= F^{2}\left( F-\frac{(\omega^{2}-q^{2})}{\alpha \omega q^{2}}  \right).
\end{equation}
Recall that all $\alpha,\beta,\omega,q>0$. For F to be real, the terms $\left(F-  \frac{(\omega^{2}-q^{2})}{\alpha \omega q^{2}} \right)$ and $\frac{\beta (q^{2}-3\omega^{2})}{6\alpha \omega}$ must have the same signs.

Denote
$ p=\frac{(\omega^2-q^2)}{\alpha\omega q^{2}}$ and introduce new variable $w(\xi)$
\begin{equation} \label{wdef}
 w(\xi)=\sqrt{\frac{p}{F(\xi)}} ~\Longrightarrow ~F(\xi)=\frac{p}{w(\xi)^{2}}. \end{equation}
 Substitution of $F(\xi)$, specified by (\ref{wdef}) into (\ref{rf1d}) gives 
\begin{equation} \label{rf1e}
 \frac{2\beta q^{2}(3\omega^{2}-q^{2})}{3(\omega^{2}-q^{2})} (w')^{2}=w^{2}-1 \quad \mbox{or} \quad \pm
\sqrt{  \frac{2\beta q^{2}(3\omega^{2}-q^{2})}{3(\omega^{2}-q^{2})}}\, \frac{d w}{d\xi}= \sqrt{w^{2}-1}.
\end{equation}
The last equation can be written as 
\begin{equation} \label{rf1f}
\frac{d w}{\sqrt{w^{2}-1}} = \pm \sqrt{\frac{3(\omega^{2}-q^{2})}{2\beta q^{2}(3\omega^{2}-q^{2})}} \, d\xi 
\end{equation}
and integrated, giving
\begin{equation} \label{rf1g}
\text{arccosh}\left(w\right)= \pm \sqrt{\frac{3(\omega^{2}-q^{2})}{2\beta q^{2}(3\omega^{2}-q^{2})}} \,(\xi +\xi_{0}). 
\end{equation}
Then
\begin{equation} \label{rf1h}
w = \sqrt{\frac{p}{F(\xi)^{2}}} = \text{cosh}\left( \pm \sqrt{\frac{3(\omega^{2}-q^{2})}{2\beta q^{2}(3\omega^{2}-q^{2})}} (\xi +\xi_{0} )\right).
\end{equation}
Finally, we obtain  ($\xi_{0}$ and $\pm$  are irrelevant)
\begin{equation} \label{rf1g1}
F(\xi)
= \frac{(\omega^{2}-q^{2})}{\alpha \omega q^{2}} \text{sech}\left( \sqrt{\frac{3(\omega^{2}-q^{2})}{2\beta q^{2}(3\omega^{2}-q^{2})}} \,\xi \right)^{2}.
\end{equation}      
Now, since $f(\xi)= \partial^{-1}_{\xi} F(\xi)$, we can express $\eta$ according to (\ref{et1}) obtaining a solitary solution for surface wave in first-order approximation for small parameters $\alpha,\beta$
\begin{align} \label{solw1}
\eta(\xi)
 & = \frac{2 (\omega^{2}-q^{2})}{\alpha (3\omega^{2}-q^{2})} \text{sech}^2\left(\sqrt{\frac{3(\omega^{2}-q^{2})}{2\beta q^{2}(3\omega^{2}-q^{2})}} \, \xi\right) \\ 
  & \hspace{2ex}  +\frac{(\omega^{2}-q^{2})^2 \left(q^2+6 \omega ^2\right)}{2
   \alpha  q^2 \omega ^2 (3\omega^{2}-q^{2})}\text{sech}^4\left(\sqrt{\frac{3(\omega^{2}-q^{2})}{2\beta q^{2}(3\omega^{2}-q^{2})}} \, \xi\right). \nonumber 
\end{align}        
We consider the case when all  $k,l,\omega>0$. By definition $\beta>0$.
Then, solitary waves (\ref{solw1}) are real when $\frac{(\omega^{2}-q^{2})}{(3\omega^{2}-q^{2})} >0$. This condition is satisfied when ~$\omega > q$.

\subsection{Examples of wave profiles} \label{exsolw}
Let us examine some examples of solitary waves of the form (\ref{solw1}).
Assume reasonable values of small parameters $\alpha,\beta\in \left(\frac{1}{10},\frac{1}{5}\right)$. The function
$\eta(\xi)$ is real when $\omega >q$. Denote $\omega = v\,q$, where $v >1$. 
Let us write the solitary wave in coordinates $x',y'$ rotated with respect to $x,y$ by the angle $\phi=\text{arctan}\frac{l}{k}$. Then $\xi=(k x +l y -\omega t) = q (x'-v t)$.
In these rotated coordinates the wave moves along $x'$, exhibiting translational symmetry for $y'$. In rotated coordinates the solitary wave  (\ref{solw1}) reads as
\begin{align} \label{solw1R}
\eta(x'-v t) 
 & =  \frac{2(v^{2}-1)}{\alpha(3v^{2}-1)}\, \text{sech}^2 \left( \sqrt{\frac{3(v^2-1)}{2(3 v^2-1)\beta}}\,(x'-v t)\right) \\ & \hspace{2ex} +\frac{\left(v^2-1\right)^2 \left(6 v^2+1\right)}{2\alpha\, v^2 (3 v^2-1)}\,  \text{sech}^4 \left(\sqrt{\frac{3(v^2-1)}{2(3 v^2-1)\beta}}\,(x'-v t)\right) \nonumber  \\ & =: \eta_{2}(x'-v t)+ \eta_{4}(x'-v t) \nonumber
\end{align} 
The solitary wave (\ref{solw1R}) is a sum of two components denoted as $\eta_{2},\eta_{4}$ which depend on $\text{sech}^2, \text{sech}^4$ functions, respectively.  

\begin{figure} 
\begin{center}
\resizebox{0.7\columnwidth}{!}{\includegraphics{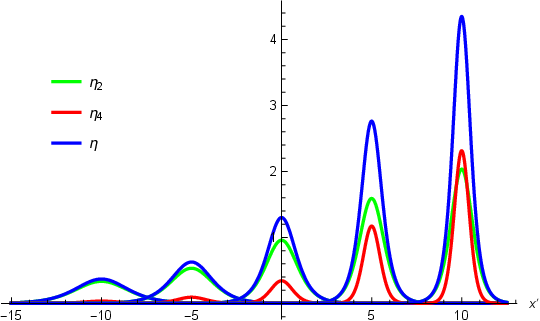}} \\ \resizebox{0.7\columnwidth}{!}{\includegraphics{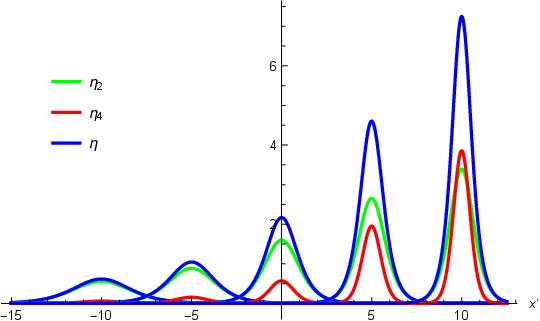}}\\
\resizebox{0.7\columnwidth}{!}{\includegraphics{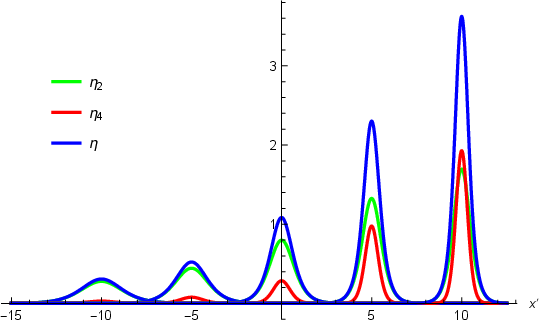}} 
\end{center}
\caption{Profiles of the solitary wave $\eta$ and their components $\eta_{2},\eta_{4}$ (\ref{solw1R}) for 
$v= 1.03, 1.05, 1.1, 1.15, 1.2 $ (from left to right). Top: $\alpha=\beta=\frac{1}{6}$. Middle: $\alpha=\frac{1}{10}, \beta=\frac{1}{5}$. Bottom: $\alpha=\frac{1}{5}, \beta=\frac{1}{10}$.
All profiles are centred at $x'=0$ but shifted artificially to avoid overlaps. For $v=1.03$, the component $\eta_{4}$ is so small that green and blue curves almost coincide.}  \label{F1A}
\end{figure}

There are two branches of real solutions. The first one contains $v>1$.
In Fig.~\ref{F1A}, we present five profiles of the solitary wave (\ref{solw1R}) for $v=1.03, 1.05, 1.1, 1.15, 1.2 $ (from left to right). Components $\eta_{1}$ and $\eta_{2}$ are displayed, as well. Each profile corresponds to $t=0$ and should be centred at $x'=0$, but to avoid overlaps, the first is shifted to the left by ten units, the second is shifted to the left by five units, and the fourth and fifth are shifted to the right by five and ten units, respectively.
The top row represents wave profiles when $\alpha=\beta=\frac{1}{6}$, the middle one is for 
$\alpha=\frac{1}{10}, \beta=\frac{1}{5}$ and the bottom one is for $\alpha=\frac{1}{5}, \beta=\frac{1}{10}$.
Generally, profiles of waves corresponding to different (but small) $\alpha,\beta$ are very similar.  
For all cases, the solution's amplitude grows rapidly with increasing $v$ and becomes unphysically large (recall that higher waves tend to break).

It is worth noting that the profile of the solitary surface wave (\ref{solw1R}) does not depend on all the parameters $k,l,\omega$ but only on their particular function, the velocity $v=\frac{\omega}{\sqrt{k^{2}+l^{2}}}$.

\begin{figure} 
\begin{center}
\resizebox{0.7\columnwidth}{!}{\includegraphics{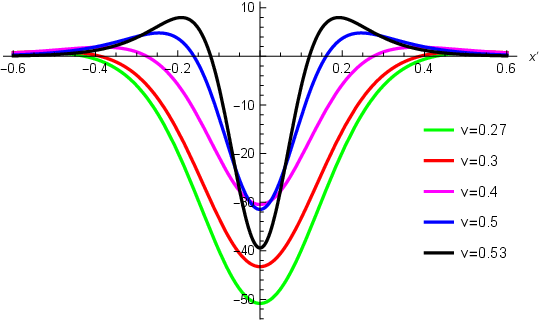}} \\ \resizebox{0.7\columnwidth}{!}{\includegraphics{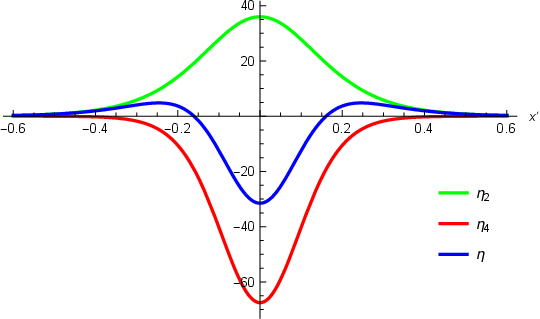}}
\end{center}
\caption{ Top: Profiles of the solitary wave (\ref{solw1R}) for $\alpha=\beta=\frac{1}{6}$ and $v= 0.27,0.3,0.4,0.5,0.53 $.  Bottom: The wave with $v=0.5$, and their components $\eta_{2}$ and $\eta_{4}$. The blue line represents the same profile, corresponding to $v=0.5$, in both pictures.} 
\label{Fv05}
\end{figure}

The second branch of real solutions contains $v \in (0,\frac{1}{\sqrt{3}}\approx 0.57735)$.
These solutions are practically unphysical because their amplitudes are very large. In Fig.~\ref{Fv05}, we present several profiles of solitary wave (\ref{solw1R}) for $v=0.27,0.3,0.4,0.5,0.53$. The smallest amplitude (still greater than 30) occurs for $v\approx 0.4-0.5$, but it grows when the velocity increases or decreases.

\subsection{Case 2, integration constants $r,s\ne 0$. Cnoidal waves}

The case when integration constants in (\ref{rf1b}) are nonzero, i.e. $r,s \ne 0$, is more complicated. 
In \cite{Ding}, Dingemans derived the periodic solutions of the KdV equation expressed by the elliptic Jacobi function $\text{cn}^{2}(B (x- vt),m)$
(equivalently by $\text{dn}^{2}(B (x- vt),m)$ or $\text{sn}^{2}(B (x- vt),m)$, where $m$ is the elliptic parameter).

\subsection{Case C=0} \label{Ceq0}

Since the equation (\ref{rf1b}) has the same form as that obtained from the KdV equation by the first integration, we can search its solution with the following ansatz
\begin{equation} \label{cn2F0}
F(\xi) = A\, \text{cn}^{2}(B \xi,m).
\end{equation}
With $F(\xi)$ in the form (\ref{cn2F0}) the equation (\ref{rf1b}) takes the following form
\begin{align} \label{r14F}
C_{6}\, \text{cn}^{6}(B \xi,m) +C_{4}\, \text{cn}^{4}(B \xi,m)+C_{2}\, \text{cn}^{2} (B \xi,m)+C_{0}  = 0,
\end{align}
where
\begin{align} \label{cc06}
C_{6} & =- A^2 \left(\frac{6 A\alpha \omega }{\beta (q^2-3\omega^{2})}+4 B^2 m\right) =0, \\   \label{cc04}
C_{4} & = \frac{2 A^2 \left(2 \beta  B^2 (2 m-1) q^4-3 q^2 \left(2 \beta  B^2 (2 m-1) \omega
   ^2+1\right)+3 \omega ^2\right)}{\beta  q^2 \left(q^2-3\omega^2\right)} =0, \\   \label{cc02}
C_{2} & = 2 A \left(-2 A B^2 (m-1)+\frac{3 r}{\beta q^2(q^2-3\omega^2)}\right) =0, \\   \label{cc00}
C_{0} & = -s=0.
\end{align}
We look for nontrivial solutions with $A\ne 0$. 
From (\ref{cc04}) and next (\ref{cc06}) we can express $B,A$ through $m$
\begin{equation} \label{B-A} 
B= \sqrt{\frac{3(q^{2}-\omega^{2})}{2 (2 m-1)q^{2}(q^2-3\omega^2)\beta}}
 , \qquad  A = \frac{ m \left(q^2-\omega^2\right)}{\alpha  (1-2 m) q^2 \omega } .
\end{equation}
Relations (\ref{B-A}) allow us to fix from (\ref{cc02}) integration constant $r$ (which does not appear in the solutions)
$$ 
r  = -\frac{ m(m-1)\left(q^2-\omega\right)^2}{\alpha  (1-2 m)^2 q^2 \omega }. $$ 
We see that there is a wide family of solutions to the equation (\ref{rf1b}) 
in the form (\ref{cn2F0}), with $A,B$ given by (\ref{B-A}), and $m\in(0,1)$. From a mathematical point of view, the conditions imposed on parameters $k,l,\omega,A,B$ of admissible solutions give much freedom. 

Now, since $f(\xi)= \partial^{-1}_{\xi} F(\xi)$, we can express $\eta$ according to (\ref{et1}) obtaining another cnoidal solution for surface wave in first-order approximation for small parameters $\alpha,\beta$
\begin{align} \label{solcn2}
\eta(\xi,m) & = A_{0} + A_{2}\, \text{cn}^{2}\left(B\,\xi,m\right) + A_{4} \, \text{cn}^{4}\left(B\,\xi,m\right) 
\end{align}
where 
\begin{align} \label{A00}
A_{0} & = -\frac{3  m (m-1) \left(q^2-\omega ^2\right)^2}{2
   \alpha (1 -2 m)^2 q^2 \left(q^2-3\omega ^2\right)}, \\   \label{A02}
A_{2} & = \frac{2 m \left(q^2-\omega ^2\right) }{ \alpha  (2 m-1) \left(q^2-3\omega ^2\right)}, \\   \label{A04}
A_{4} & = -\frac{m^2 \left(q^2-\omega ^2\right)^2 \left(q^2+6\omega^2 \right)}
{2 \alpha  (1-2 m)^2 q^2 \omega^2\left(q^2-3\omega ^2\right)}. 
\end{align}

As we stressed in \cite{IKRR,RK2023NoDy}, periodic solutions must satisfy mass (volume) conservation condition. This means that the volumes of water uplifted and downlifted from the equilibrium level must be equal to each other over an interval equal to the wavelength. 
The equation that expresses this condition is the following
\begin{equation} \label{genVCC}
\int_{0}^{L} \left(\eta(\xi,m)+D\right) \, d\xi=0,
\end{equation}
where $L$ is the wavelength (space period) of $\eta$ and $D$ is a constant assuring volume conservation.
For $\text{cn}^{2}(B\xi,m)$, $\text{cn}^{4}(B\xi,m)$ the period is $L=\frac{2 K(m)}{B}$, where $K(m)$ is is the complete elliptic integral of the first kind. Because 
\begin{align} \label{Icn2}
I_{2} & =\int_{0}^{2K(m)/B} \text{cn}^{2}(B\xi,m)\, d\xi  = \frac{2 E(m)+(m-1) K(m)}{B m} \quad \mbox{and} \\  \label{Icn4}
I_{4} & =\int_{0}^{2K(m)/B} \text{cn}^{4}(B\xi,m)\, d\xi  = \frac{(8 m-4) E(m)+2(2-5m+3m^{2}) K(m)}{3 B m^{2}},
\end{align}
where $E(m)$ is the complete elliptic integral, the volume conservation condition (\ref{genVCC}) requires
$$ \int_{0}^{\frac{2K(m)}{B}}\!\! \!\left( A_{0}+D + A_{2}\, \text{cn}^{2}\left(B\,\xi,m\right) + A_{4} \, \text{cn}^{4}\left(B\,\xi,m\right)\right) \, d\xi =( A_{0}+D)\frac{2K(m)}{B} + A_{2} I_{2}+A_{4}I_{4}=0.$$ 
Therefore 
\begin{align}  \label{A0D}
A_{00}=A_{0}+D & = -\frac{( A_{2} I_{2}+A_{4}I_{4}) B}{2 K(m)}.
\end{align}

So, we finally obtained the first-order (in $\alpha,\beta$) approximation for traveling 
periodic surface waves determined by the (2+1)-dimensional Boussinesq equations (\ref{BR7})-(\ref{BR8}) as
\begin{align} \label{solFcn24}
\eta(\xi,m) & = A_{00} + A_{2}\, \text{cn}^{2}\left(B\,\xi,m\right) + A_{4} \, \text{cn}^{4}\left(B\,\xi,m\right),
\end{align}
where the constant $A_{00}$ assures volume conservation condition. 

\subsection{Examples of cnoidal waves, C=0 } \label{excn}
Let us examine some examples of solitary waves of the form (\ref{solFcn24}).
Assume reasonable values of small parameters $\alpha=\beta=\frac{1}{6}$.  
Let us write the solitary wave in coordinates $x',y'$ rotated with respect to $x,y$ by the angle $\phi=\text{arctan}\frac{l}{k}$. Denote $\omega = v\,q$. 
Then $\xi=(k x +l y -\omega t) = q (x'-v t)$ and the function $\eta(\xi,m)$ depends only 
$x', v, t$ and $m $.
In these rotated coordinates the wave moves along $x'$, exhibiting translational symmetry for $y'$. In rotated coordinates the solitary wave  (\ref{solFcn24}) reads as
\begin{align} \label{etaP}
\eta(x',t,v,m) & = A_{00} +A_{02}\, \text{cn}^{2} \left(3\sqrt{\frac{(1-v^{2})}{(2m-1)(1-3v^{2})}}\,(x'-v t),m\!\right) \nonumber\\& \hspace{6ex}+ A_{04}\, \text{cn}^{4}\left(3\sqrt{\frac{(1-v^{2})}{(2m-1)(1-3v^{2})}}\,(x'-v t),m\!\right) \\ & = A_{00} + \eta_{2}(x',m) + \eta_{4}(x',m) , \nonumber
\end{align}
where 
\begin{align} \label{A00a}
 A_{00} & = -\frac{\left(v^2-1\right)}{(1-2 m)^2 v^2 \left(3 v^2-1\right)} 
 \Big[ 2 (2 m-1) \left(6 v^4+v^2-1\right)\frac{E(m)}{K(m)} \\ & \hspace{26ex} +
 (m-1) \left(3 m \left(6 v^4+3 v^2-1\right)-2 \left(6 v^4+v^2-1\right)\right)\Big] \nonumber  \\  \label{A024}
A_{02} & = \frac{12 m \left(v^2-1\right)}{(2 m-1) \left(3 v^2-1\right)} , \quad 
A_{04} = \frac{3 m^2 \left(v^2-1\right)^2 \left(6 v^2+1\right)}
{(1-2 m)^2 v^2 \left(3 v^2-1\right)}.
\end{align} 

\begin{figure} 
\resizebox{0.733\columnwidth}{!}{\includegraphics{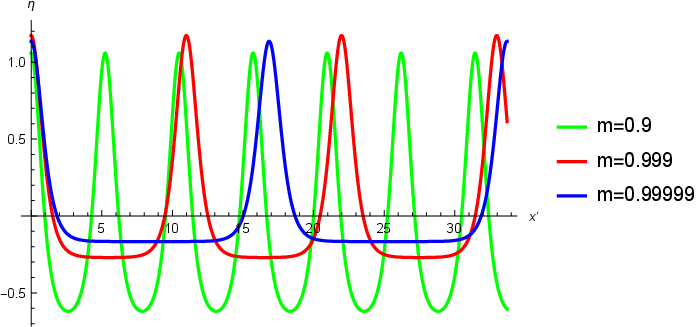}}\\   \resizebox{0.70\columnwidth}{!}{\includegraphics{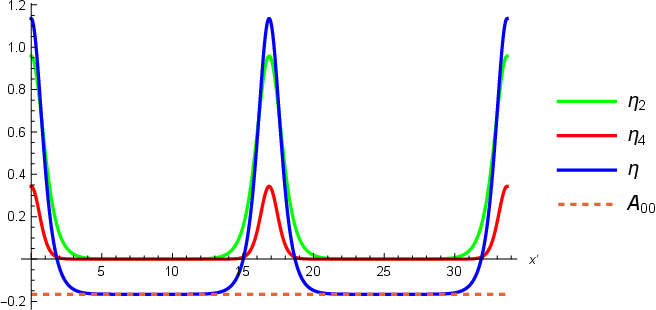}} 
\caption{Top: Profiles of the cnoidal wave (\ref{etaP}) for $v=1.1$  and three values of $m= 0.9,0.999, 0.99999$. Bottom: The profile of the cnoidal wave (\ref{etaP}) for $v=1.1$, $m=0.99999$  and its $\eta_{2}$, $\eta_{2}$ and $A_{00}$ components. The blue line represents the $\eta$ in both pictures.
}  \label{f1A}
\end{figure}

The cnoidal wave  $\eta(\xi,m)$ given by (\ref{solFcn24}) is real when $B\in\mathbb{R}$. In rotated coordinates this condition becomes $\frac{(1-v^{2})}{(2m-1)(1-3v^{2})}>0$. Therefore there are two branches of admissible solutions
\begin{itemize} 
\item $m\in (\frac{1}{2},1)$~ implying ~$v>1$.
\item $m\in (0,\frac{1}{2})$~ implying ~$\frac{1}{\sqrt{3}}< v<1$.
\end{itemize}

Consider the branch with $m\in (\frac{1}{2},1)$. Let us begin with $m$ values close to 1.  
Three examples of profiles of the waves (\ref{etaP}) belonging to the first branch are presented in the top part of Fig.~\ref{f1A}. In these cases, $m$ is close to 1, and $v$ slightly exceeds 1.
For the same $m$ values, when $v$ increases, the amplitudes increase to unphysical values.
When ~$m\to 1$, the space period increases to infinity, and the wave profile tends to solitary wave. 
In the bottom part of Fig.~\ref{f1A}, we show the decomposition of the wave (\ref{etaP}), for $v=1.1$, $v=0.999$, into its components $\eta_{2},\eta_{4}$ and $A_{00}.$.

\begin{figure} 
\begin{center}
\resizebox{0.7\columnwidth}{!}{\includegraphics{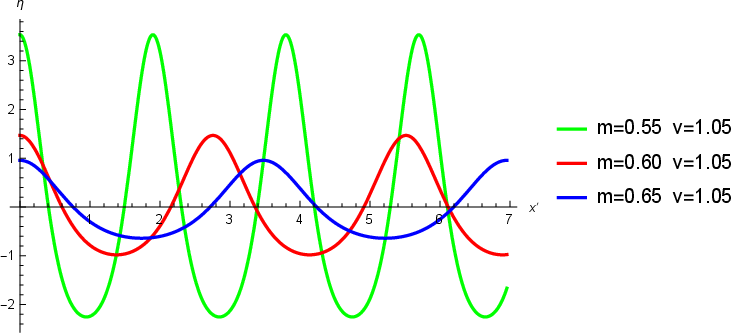}} \\
\resizebox{0.7\columnwidth}{!}{\includegraphics{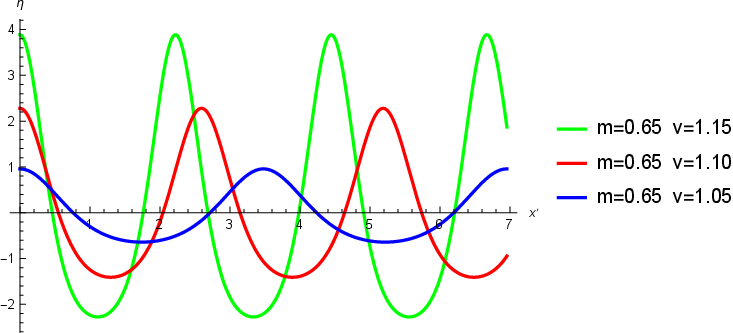}} \\ \hspace{-7ex}
\resizebox{0.645\columnwidth}{!}{\includegraphics{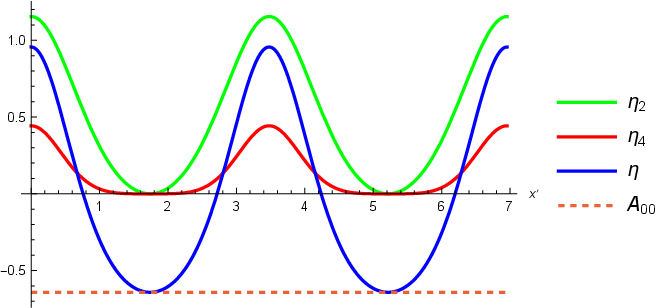}}
\end{center}
\caption{Top: Profiles of the cnoidal wave (\ref{etaP}) with $v=1.05$, and three values of $m$, close to $\frac{1}{2}$. Middle:  Profiles of the cnoidal wave (\ref{etaP}) for  $m=0.65$, and $v$ slightly larger than $1$. Bottom: The wave with $v=1.05$, $m=0.65$, and its components $\eta_{2}, \eta_{4}, A_{00}$. The blue curve represents the same wave in all pictures (note different vertical scales). }  \label{f1E}
\end{figure}

When $m$ is reduced to about $\frac{1}{2}$, cnoidal solutions (\ref{etaP}) become similar to cosinusoidal (sinusoidal) waves. This is illustrated in Fig.~\ref{f1E} (top), where $m=0.55, 0.60, 0.65$ and $v=1.05$, and in Fig.~\ref{f1E} (middle), where $m=0.65$ and $v=1.05, 1.10, 1.15$. However, the wave amplitude becomes unphysically large when $m\to \frac{1}{2}$. Fig.~\ref{f1E} (bottom) displays the 
components $\eta_{2}, \eta_{4}, A_{00}$ of $\eta$ for $m=0.65$ and $v=1.05$.

\begin{figure}  \begin{center}
\resizebox{0.7\columnwidth}{!}{\includegraphics{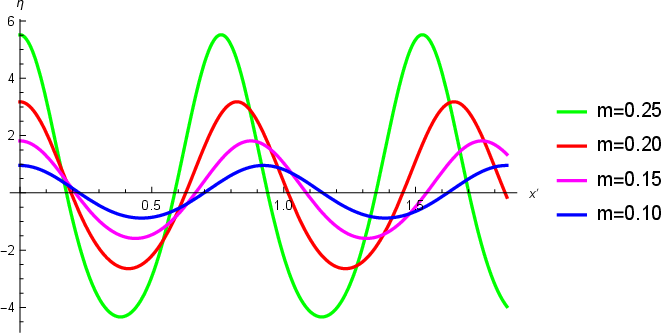}} \\\resizebox{0.723\columnwidth}{!}{\includegraphics{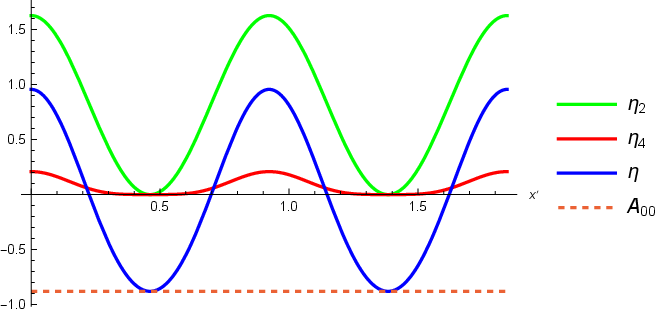}}  
\end{center}
\caption{Top: Profiles of the cnoidal wave (\ref{etaP}) for $v=0.7$, and four values of $m=0.25, 0.20, 0.15, 0.10$.  Bottom: Profile of the cnoidal wave (\ref{etaP}) for $m=0.10, v=0.7$,   and its components $\eta_{2}, \eta_{4}, A_{00}$. The blue curve represents the same wave (note different vertical scales).}  \label{f1D} \end{figure}

Solutions belonging to a branch $m\in (0,\frac{1}{2})$ are less interesting. 
In this branch admissble $v\in(\frac{1}{\sqrt{3}},1)$.
In Fig.~\ref{f1D}, we show some examples of profiles of such solutions with amplitudes not much different from unity. When $m<0.3$, all waves given by (\ref{etaP}) have shapes very similar to sinusoidal (cosinusoidal) waves.

\subsection{Case $C \ne 0$} \label{Cneq0}

The case when integration constants in (\ref{rf1b}) are nonzero, i.e. $r,s \ne 0$, is more complicated. In \cite{Ding}, Dingemans demonstrated the existence of periodic solutions to the KdV equation in the form of the Jacobi elliptic functions, for instance
\begin{equation} \label{cn2}
\eta(x,t) = A\, \text{cn}^{2}[B (x-v t),m] + C,
\end{equation}
where constants $A,B,C$ and the elliptic parameter $m$ are determined by the coefficients of the equation~(\ref{rf1b}). (Alternatively, the elliptic modulus $k$ may be used for the Jacobi elliptic functions, where $k^{2}=m$.) 
Thanks to the relationship between the Jacobi elliptic functions, periodic solutions (\ref{cn2}) can be equivalently expressed using sn$^2(B \xi,m)$ or dn$^2(B \xi,m)$ functions.

Since equation (\ref{rf1b}) has the same form as that considered by Dingemans in \cite{Ding}, we can search for its solution by taking the ansatz
\begin{equation} \label{cn2F}
F(\xi) = A\, \text{cn}^{2}(B \xi,m) + C.
\end{equation}
With $F(\xi)$ in the form (\ref{cn2F}) equation (\ref{rf1b}) takes the following form
\begin{align} \label{r14Fa}
C_{6}\, \text{cn}^{6}(B \xi,m) +C_{4}\, \text{cn}^{4}(B \xi,m)+C_{2}\, \text{cn}^{2} (B \xi,m)+C_{0}  = 0,
\end{align}
where
\begin{align} \label{c06}
C_{6} & = A^2 \left(\frac{3 \alpha  A \omega }{\beta  q^2}-4 B^2 m\right) =0, \\   \label{c04}
C_{4} & = \frac{A^2 \left(4 \beta  B^2 (2 m-1) q^4+q^2 (9 \alpha  C \omega
   +3)-3 \omega ^2\right)}{\beta  q^4} =0, \\   \label{c02}
C_{2} & = \frac{A \left(-4 A \beta  B^2 (m-1) q^4+9 \alpha  C^2 q^2 \omega
   +6 C \left(q^2-\omega ^2\right)-3 r\right)}{\beta  q^4} =0, \\   \label{c00}
C_{0} & =  \frac{3 \alpha  C^3 q^2 \omega +3 C^2 \left(q^2-\omega
   ^2\right)-3 C r- \beta  q^4 s}{\beta  q^4}=0.
\end{align}
Equation (\ref{r14Fa}) is satisfied when all $C_{i}$ coefficient vanish.
From (\ref{c06}) and  (\ref{c04}) we can express $B,C$ through $A$ and $m$
\begin{equation} \label{B-C-A} 
 B= \frac{1}{2 q} \sqrt{\frac{3\alpha \omega A}{\beta m}}, \qquad 
 C = \frac{\alpha A q^2 \omega(1-2m) -m (q^2- \omega ^2)}
 {3 \alpha  m q^2 \omega }.
\end{equation}
Relations (\ref{B-C-A}) allow us to fix from (\ref{c02})-(\ref{c00}) integration constants $r,s$, which, however, do not appear in the function (\ref{cn2F})
\begin{align} \label{rr2}
r & =\frac{1}{3}\left( \frac{2 \omega }{\alpha }+q^2 \left(\frac{\alpha  A^2
   \left(m^2-m+1\right) \omega }{ m^2}-\frac{1}{\alpha \omega }\right)-\frac{\omega ^3}{\alpha  q^2} \right)\\ \label{ss0}
s & = -\frac{\left(\alpha  A (2 m-1) q^2 \omega +m (q^2-\omega^2)\right) }
   {9 \alpha ^2 \beta  m^3 q^8 \omega ^2} \\ & \hspace{3ex} \times
   \left(\alpha  A (m-2) q^2 \omega -m (q^2- \omega^2)\right) 
   \left(\alpha  A (m+1) q^2 (\omega -m (q^2- \omega^2)\right) \nonumber
\end{align} 

Now, since $f(\xi)= \partial^{-1}_{\xi} F(\xi)$, we can express $\eta$ according to (\ref{et1}) obtaining another cnoidal solution for surface wave in first-order approximation for small parameters $\alpha,\beta$
\begin{align} \label{scn2C}
\eta(\xi,m) & =  A_{0} + A_{2}\, \text{cn}^{2}\left(\ \frac{1}{2 q} \sqrt{\frac{3\alpha  \omega A}{\beta m}}\,\xi,m\right) + A_{4} \, \text{cn}^{4}\left( \frac{1}{2 q} \sqrt{\frac{3\alpha \omega A}{\beta m}}\,\xi,m\right),
\end{align} 
where 
\begin{align} \label{aC0}
A_{0} & = \frac{1}{36} \! \left( \!-\frac{\alpha  A^2 \left(2 (1 \!- \!2 m)^2 q^2 \!- \!27
   (m \!- \!1) m \omega ^2\right)}{m^2} \!- \!\frac{4 A (2 m \!-1 \!) \left(q^2 \!+ \!2
   \omega ^2\right)}{m \omega } \!+ \!\frac{\frac{10 \omega ^2}{q^2} \!- \!\frac{2 q^2}{\omega ^2} \!- \!8}{\alpha } \!\right), \\   \label{aC2}
A_{2} & =  \frac{A \left(9 \alpha  A (1-2 m) \omega ^3+2 \alpha  A (2 m-1)
   q^2 \omega +2 m q^2+4 m \omega ^2\right)}{6 m \omega }, \\ \label{aC4}
A_{4} & = \frac{1}{4} \alpha  A^2 \left(9 \omega ^2-2 q^2\right)  
\end{align}

Finally, we need to ensure that the mass (volume) of the solution (\ref{scn2C})  is preserved.
Proceeding in the same way as in section 3.1 (equations (\ref{genVCC})-(\ref{A0D})), we obtain
\begin{align} \label{scn2CV}
\eta(\xi,m) & =  A_{0V} + A_{2}\, \text{cn}^{2}\left(\ \frac{1}{2 q} \sqrt{\frac{3\alpha  \omega A}{\beta m}}\,\xi,m\right) + A_{4} \, \text{cn}^{4}\left( \frac{1}{2 q} \sqrt{\frac{3\alpha \omega A}{\beta m}}\,\xi,m\right),
\end{align}
where
\begin{align} \label{A0vc}
 A_{00} & = - A_{2}\left(2 \frac{E(m)}{K(m)}+(m-1) \right) - A_{4} \left( (8 m-4)  \frac{E(m)}{K(m)}+2(2-5m+3m^{2})\right) 
\end{align}
ensures that the volume of fluid raised and lowered relative to the undisturbed level is equal.

\subsection{Examples of cnoidal waves, case $C\ne 0$ } \label{cn24Cn0}

Let us examine some examples of solitary waves of the form (\ref{scn2CV}).
Assume reasonable values of small parameters $\alpha=\beta=\frac{1}{6}$.  
Let us write the solitary wave in coordinates $x',y'$ rotated relative to $x,y$ coordinates by the angle $\phi=\text{arctan}\frac{l}{k}$. Denote $\omega = v\,q$. 
Then $\xi=(k x +l y -\omega t) = q (x'-v t)$ and the function $\eta(\xi,m)$ depends on 
$x', v, q, t$ and $m $. 
In rotated coordinates, the cnoidal wave  (\ref{scn2CV}) reads as
\begin{equation} \label{scn2CVr}
\eta(x',t,v,q,m)  =  A_{00} + A_{2}\, \text{cn}^{2}\left[ \sqrt{\frac{3 v q A}{4 m}}\,(x'-vt),m\right] + A_{4} \, \text{cn}^{4}\left[ \sqrt{\frac{3 v q A}{4m}}\,(x'-vt),m\right],
\end{equation}
where 
\begin{align} \label{acn2}
A_{2} & = \frac{A q \left(2 m \left(A q \left(2-9 v^2\right) v+12
   v^2+6\right)+A q v \left(9 v^2-2\right)\right)}{36 m v}, \\ \label{acn4}
A_{4} & = \frac{1}{24} A^2 q^2 \left(9 v^2-2\right), \\ \label{a0v}
A_{00} & = -\frac{A q }{72 m^2 v}
 \Big((m-1) \left(m \left(A q \left(9 v^2-2\right) v+24 v^2+12\right) 
   +A q v \left(2-9 v^2\right)\right) \nonumber \\ & \hspace{12ex}
+24 m\left(2 v^2+1\right) \frac{E(m)}{K(m)}\Big).
\end{align}

From a mathematical point of view, the parameters determining the solution (\ref{scn2CVr})  are not restricted; $A, v, q$ and $ m\in (0,1)$ can be arbitrary. However, physical solutions are limited to small amplitude waves (in scaled variables, this means an amplitude close to unity). Therefore, in the examples below, we only show such waves.

\begin{figure} 
\begin{center}
\resizebox{0.7\columnwidth}{!}{\includegraphics{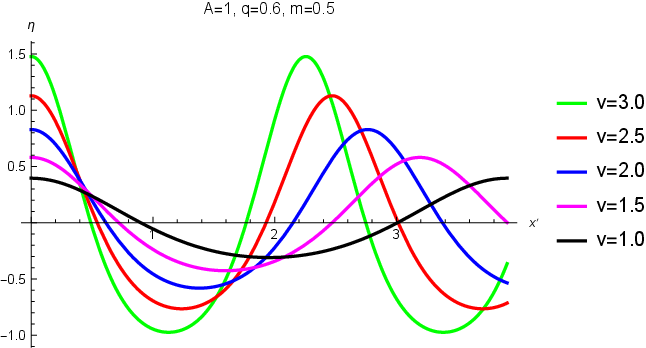}} \\ 
\resizebox{0.68\columnwidth}{!}{\includegraphics{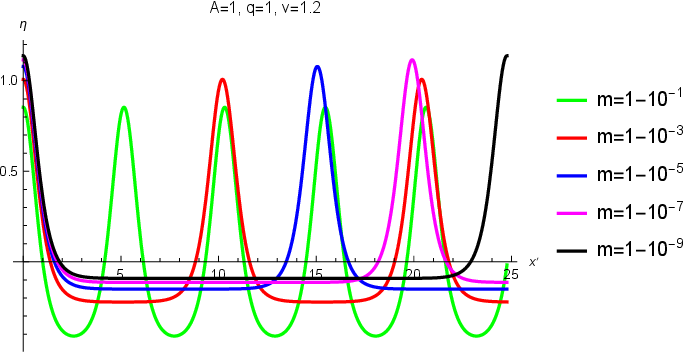}} \\
\resizebox{0.7\columnwidth}{!}{\includegraphics{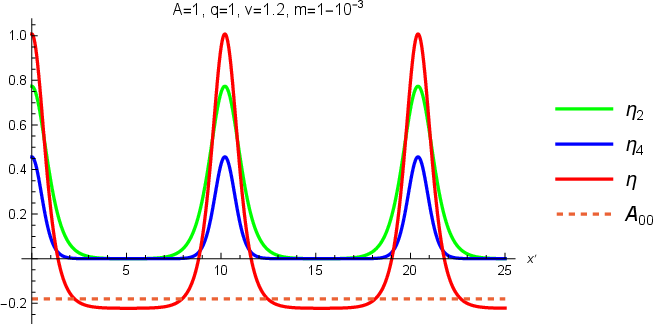}}
\end{center}
\caption{Top: Profiles of the cnoidal wave (\ref{scn2CVr}) for $m=0.5$,  and five values of $v$. Middle: Profiles of the cnoidal wave (\ref{scn2CVr}) for $v=0.98$, and five values of $m$, close to 1. Bottom: The wave with $m=0.999$ from the middle plot and its components $\eta_{2}, \eta_{4}, A_{00}$. The red curve represents the same wave in the middle and bottom plots.
}  \label{fperCne0}
\end{figure}
In Fig.~\ref{fperCne0}, we show several examples of profiles of waves given by the equation (\ref{scn2CV}) transformed to rotated coordinates when $\alpha=\beta=\frac{1}{6}$. The top plot displays profiles of waves with four different velocities $v=3, 2.5, 2, 1.5, 1$ with fixed parameters $A=1, ~q=0.6, ~m=0.5$. The middle plot displays profiles with fixed $A=1, ~q=1, ~v=1.2$ and different $m\to 1$.  One of these profiles, corresponding to $m=0.999$, is in the bottom plot decomposed into its components $\eta_{2}, \eta_{4}$ and $A_{00}$.

\subsection{Superposition solutions} \label{sups}

In 2013, Khare and Saxena \cite{KhSa} showed for the first time that periodic solutions other than the usual cnoidal solutions exist for the KdV equation. 
These solutions have the following form
\begin{equation} \label{ssup}
\eta(x,t) = A\left(\text{dn}^{2}[B(x-vt),m] \pm\sqrt{m}\,\text{cn}[B(x-vt),m]\,\text{dn}[B(x-vt),m] \right).
\end{equation}
Since the function (\ref{ssup}) is the sum of two different periodic functions, this type of solution gains the name \emph{superposition solution}. Later, it appeared that similar solutions exist for other nonlinear wave equations, see, e.g.\ \cite{RKI,KRBook,KhSa14,KhSa15}. 

Since the equation (\ref{rf1b}) has a form analogous to that of the KdV equation, we can search for its solution in the form of an analogous superposition 
\begin{equation} \label{ssupC}
F(\xi) = A\left(\text{dn}^{2}[B\xi,m] \pm\sqrt{m}\,\text{cn}[B\xi,m]\,\text{dn}[B\xi,m] \right) +C.
\end{equation}
Inserting (\ref{ssupC}) into (\ref{rf1b})  gives the following equation
\begin{align} \label{r14S}
& -\frac{1}{\beta  q^2 \left(q^2-3 \omega ^2\right)}
\Big( C_{00}+C_{11}\text{cn}[B\xi,m]\,\text{dn}[B\xi,m]+C_{31}\text{cn}^{3}[B\xi,m]\,\text{dn}[B\xi,m] \\ &\hspace{2ex} +C_{51}\text{cn}^{5}[B\xi,m]\,\text{dn}[B\xi,m] 
+C_{20}\text{cn}^{2}[B\xi,m]+C_{4}\text{cn}^{4}[B\xi,m]+C_{60}\text{cn}^{6}[B\xi,m]\Big) =0, \nonumber  \end{align}
where the coefficients $C_{ij}$ are (for equation (\ref{r14S}) to be satisfied, all $C_{ij}$ must vanish)
\begin{align} \label{C00}
C_{00} = & -6 \alpha  A^3 (m-1)^3 q^2 \omega -A^2 (m-1)^2 \left(-3 q^2 \left(\beta  B^2 m \omega ^2+6 \alpha  C \omega +2\right)+\beta  B^2 m q^4+6 \omega ^2\right)
\nonumber  \\ &  +6 A (m-1) \left(-3 \alpha  C^2 q^2 \omega -2 C q^2+2 C \omega^2 +r\right)+6 \alpha  C^3 q^2 \omega +6 C^2 \left(q^2-\omega ^2\right) \nonumber\\ 
   & -6 C r+\beta  q^2 s \left(q^2-3 \omega^2\right) =0, \\  \label{C11}
C_{11} = & 2 A \sqrt{m} \left(9 \alpha  A^2 (m-1)^2 q^2 \omega +2 A (m-1) \left(-3 q^2 \left(\beta  B^2 m \omega ^2+3 \alpha  C \omega +1\right)
 \right.\right. \nonumber \\ & \left.\left.
   +\beta  B^2 m q^4+3 \omega ^2\right)+9 \alpha  C^2 q^2 \omega +6 C \left(q^2-\omega ^2\right)-3  r\right) =0, \\ \label{C20}
C_{20} = & A m \left(36 \alpha  A^2 (m-1)^2 q^2 \omega +A (m-1) \left(-3 q^2 \left(\beta  B^2 (9 m-1) \omega ^2+18 \alpha  C
   \omega +6\right)  \right.\right. \nonumber \\ & \left.\left.
   +\beta  B^2 (9 m-1) q^4+18 \omega ^2\right)-6 \left(-3 \alpha  C^2 q^2 \omega -2 C q^2+2 C \omega
   ^2+r\right)\right)
 =0, \\  \label{C31} 
C_{31} = & 2 A^2 m^{3/2} \left(3 q^2 \left(-7 \alpha  A (m-1) \omega -2 \beta  B^2 \omega ^2+6 \beta  B^2 m \omega ^2+6 \alpha C \omega +2\right) 
 \right. \nonumber \\ & \left.
   +2 \beta  B^2 (1-3 m) q^4-6 \omega ^2\right) =0, \\  \label{C40} 
C_{40} = & 2 A^2 m^2 \left(3 q^2 \left(-9 \alpha  A (m-1) \omega -4 \beta  B^2  \omega ^2+8 \beta  B^2 m \omega ^2+6 \alpha  C \omega +2\right)  
\right. \nonumber \\ & \left.
   +4 \beta  B^2 (1-2 m) q^4-6 \omega ^2\right) =0, \\ \label{C51}
C_{51} = & 8 A^2 m^{5/2} q^2 \left(3 \alpha  A \omega +\beta  B^2 \left(q^2-3 \omega ^2\right)\right) =0,  \\ \label{C60} 
C_{60} = & 8 A^2 m^3 q^2 \left(3 \alpha  A \omega +\beta  B^2 \left(q^2-3 \omega ^2\right)\right) =0.
\end{align}
We are only interested in non-trivial solutions with $A\ne0$. We see that $C_{51}$ and $C_{60}$ are linearly dependent, so from both (\ref{C51}) and (\ref{C60}) we get $B$ as
\begin{equation} \label{bbB}
 B=\sqrt{\frac{3 \alpha  \omega  A}{ \beta (3 \omega^{2}-q^{2})}}.
\end{equation} 
With this $B$, both equations (\ref{C31}) and (\ref{C40}) give the same dependence $C(A,m)$
\begin{equation} \label{aaA}
C = \frac{m-5}{6}A + \frac{\omega^{2}-q^{2}}{3 \alpha\omega q^{2} }.
\end{equation}
Inserting these $B$ and $C$ into equations (\ref{C11}) and (\ref{C20}) one obtains the same equation
\begin{align} \label{11:20C}
&  \frac{q^4 \left(\alpha ^2 A^2 \left(m^2+14 m+1\right) \omega ^2-4\right)+4
   q^2 \omega  (2 \omega -3 \alpha  r)-4 \omega ^4}{2 \alpha  q^2 \omega } =0,
\end{align}
which allows us to determine $r(A,m)$
$$ r= \frac{q^4 \left(\alpha ^2 A^2 \left(m^2+14 m+1\right) \omega ^2-4\right)+8
   q^2 \omega ^2-4 \omega ^4}{12 \alpha  q^2 \omega }. $$ Next, inserting $B,C,r$ into (\ref{C00}) allows us to fix s(A,m) as
\begin{align} \nonumber
s & = \frac{1}{18 \alpha ^2 q^4 \omega ^2 \left(\beta 
   q^4-3 \beta  q^2 \omega ^2\right)} \Big(3 q^4 \omega ^2 \left(\alpha ^2 A^2 \left(m^2+14 m+1\right) \omega ^2-4\right) \\ & 
 +q^6 \left(\alpha ^3 A^3 \left(m^3-33 m^2-33 m+1\right)
   \omega ^3-3 \alpha ^2 A^2 \left(m^2+14 m+1\right) \omega ^2+4\right)+12
   q^2 \omega ^4-4 \omega ^6\Big) \nonumber
\end{align}
Indeed, we found that $F(\xi)$ in superposition form (\ref{ssupC}) satisfy equation (\ref{rf1b}). There exists a wide family of such solutions to equation (\ref{rf1b}) depending on two parameters $A$ and $m\in (0,1)$.
 
Next, with $F(\xi)$ in the form (\ref{ssupC}) (and then $f(\xi)=\partial^{-1}_{\xi} F(\xi)$) equation (\ref{et1}) takes the following form
\begin{align} \label{supEta}
\eta & = A_{00} +A_{20}\, \text{dn}^{2}+A_{11}\, \text{dn}\,\text{cn}+ A_{40} \text{dn}^{4}+ A_{31}\, \text{dn}^{3}\,\text{cn},\end{align}
where the argument of each Jacobi elliptic function is $(B \xi,m)$, and 
\begin{align}  \label{A20s}
A_{20} = & \frac{A \left(q^6 (1-2 \alpha  A (m-2) \omega )-q^4 \omega ^2 (12 \alpha  A (m-2) \omega +1)-6 q^2 \omega ^4\right)}{3 q^2 \omega  \left(q^2-3 \omega
   ^2\right)}  \\ \label{A11s}
A_{11} = & -\frac{A \sqrt{m} \left(q^4 (\alpha  A (m-5) \omega -2)+2 q^2 \omega ^2 (3 \alpha  A (m-5) \omega +1)+12 \omega ^4\right)}{6 \omega  \left(q^2-3
   \omega ^2\right)},  \\ \label{A40s}
A_{40} = &\frac{\alpha  A^2 q^2 \left(q^2+6 \omega ^2\right)}{3
   \omega ^2-q^2 },  \\ \label{A31s}
A_{31} = & \frac{\alpha  A^2 q^2 \sqrt{m} \left(q^2+6 \omega ^2\right)}{3
   \omega ^2-q^2 } .
\end{align}
Imposing the volume-preserving condition, analogous to (\ref{genVCC}), on the solution (\ref{supEta}), we obtain the following value of the constant $A_{00}$
\begin{align} \label{A00sV}
A_{00} & = -\frac{A \left(\alpha  A (m-1) q^2 \omega \left(q^2+6 \omega
   ^2\right) K(m) + \left(-q^4+q^2 \omega ^2+6 \omega ^4\right)E(m) \right)}{3
   \omega  \left(3 \omega ^2-q^2\right) K(m)}. \end{align}
Formula (\ref{supEta}) with coefficients (\ref{A20s})-(\ref{A00sV}) gives the form of waves that are superposition solutions of the Boussinesq equations (\ref{BR7a})-(\ref{BR8a}).

\subsection{Realistic examples} \label{exSup}

\begin{figure}[t]
\begin{center}
\resizebox{0.7\columnwidth}{!}{\includegraphics{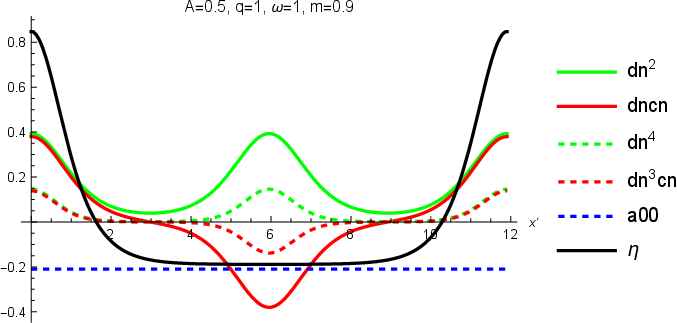}}  \\ 
\resizebox{0.7\columnwidth}{!}{\includegraphics{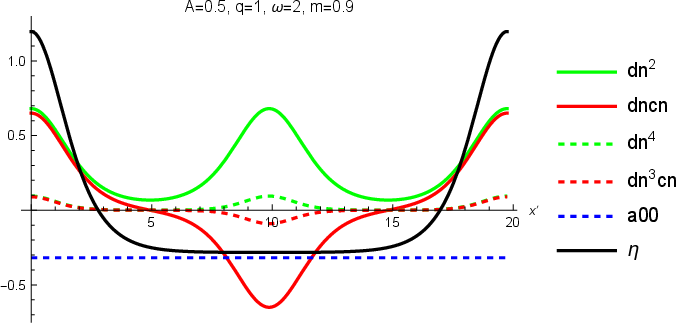}}   
\end{center}
\caption{Top: The profile of the wave (\ref{supEta}) for $A=1, ~q=1, ~\omega=1,  ~m=0.9$ and its components. Bottom: The same for $A=\frac{1}{2}, ~q=1, ~\omega=2, ~m=0.9$. The black curve labelled by $\eta$ represents the sum of all components.}  \label{f1sup}
\end{figure}

As previously, we will discuss examples of solutions (\ref{supEta}), originating from superposition solutions to equation (\ref{rf1b}), 
expressed in rotated coordinates (see subsection \ref{exsolw} for their definition). In these coordinates $k x+l y-\omega t= q(x'-v t)$, where $q=\sqrt{k^{2}+l^{2}}$, $v=\frac{\omega}{q}$.  Let us consider realistic values of small parameters $\alpha=\beta=\frac{1}{6}$. 
Then, the argument of the Jacobi elliptic functions is
\begin{equation} \label{argS}
 (B \xi,m) = \left(\sqrt{\frac{3 \alpha \omega A}{\beta (3 \omega^{2}-q^{2})}}\,(k x+l y-\omega t),m\right) = \left(\sqrt{\frac{ 3q v A }{3 v^{2}-1 } }\, (x'-v t),m\right)  
\end{equation}
In this case, coefficients given by (\ref{A20s})-(\ref{A00sV}) and the function (\ref{supEta}) depend on four parameters $A,q,\omega,m$ and introduction $\omega=q v$ does not reduce this number. \\ \indent
Mathematically, the parameters $A,q,\omega,m$ defining the periodic wave (\ref{supEta}) can be arbitrary (with $m\in(0,1)$ and $A>0, ~\omega>\frac{q}{\sqrt{3}}$ or $A<0, ~\omega<\frac{q}{\sqrt{3}}$), since the conditions (\ref{A20s})-(\ref{A00sV}) do not require any constrains. \marginpar{\textcolor{red}{Sprawdzić $A<0$.}}
Conditions $A>0, ~\omega>\frac{q}{\sqrt{3}}$ or $A<0, ~\omega<\frac{q}{\sqrt{3}}$ ensure the real value of the arguments of  Jacobi elliptic functions.
However, according to the general theory, only small amplitude waves can make physical sense, so the amplitude should be close to 1 in scaled variables.

Below, we present some examples of profiles of waves (\ref{supEta}) with small amplitudes.

Begin with waves having $A>0, ~\omega>\frac{q}{\sqrt{3}}$.
In Fig.~\ref{f1sup}, two cases of wave profiles are displayed, with $A=\frac{1}{2}, q=1, m=0.9$ and $\omega=1$ (top) and $\omega=2$ (bottom). In addition to the full solution $\eta$ of the function (\ref{supEta}) represented by the black curve, its components denoted by $\text{dn}^{2}, \text{dn}\text{cn}, \text{dn}^{4}, \text{dn}^{3}\text{cn}$ and $A_{00}$ are also shown. 

\begin{figure}[t]
\begin{center}
\resizebox{0.7\columnwidth}{!}{\includegraphics{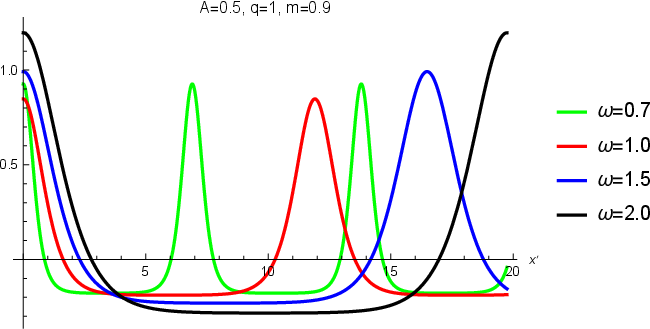}} \\ 
\resizebox{0.7\columnwidth}{!}{\includegraphics{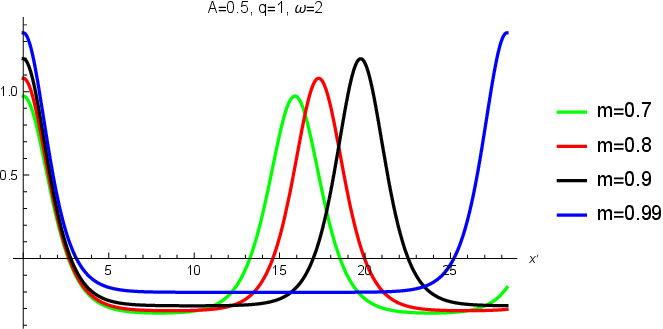}}   
\end{center}
\caption{Top: The profile of the wave (\ref{supEta}) for $A=\frac{1}{2}, q=1, m=0.9$ and four values of $\omega=0.7, 1, 1.5, 2$. Bottom: The same for $A=\frac{1}{2}, \omega=2$ and four values of $m=0.7,0.8,0.9,0.99$. Note that the black curve represents the same wave.}  \label{f2sup}
\end{figure}
In Fig.~\ref{f2sup} (top), we show profiles of the wave (\ref{supEta}) with $A=\frac{1}{2}, q=1, m=0.9$ and four values of $\omega=1, 1.5, 2, 3$. We see faster waves (greater $\omega$) have greater wavelengths. In Fig.~\ref{f2sup} (bottom), we present the waves with the same 
$A=\frac{1}{2}, q=1, \omega=3$ but four different elliptic parameters $m=0.3,0.5,0.7,0.9$. These plots show that waves with greater $m$ have greater wavelength.
  
\begin{figure}[tbh]
\begin{center}
\resizebox{0.7\columnwidth}{!}{\includegraphics{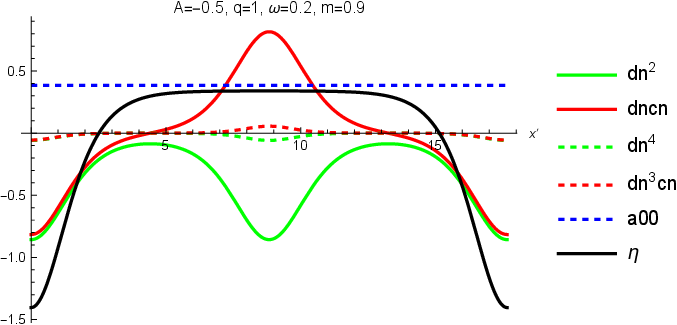}} \\ 
\resizebox{0.7\columnwidth}{!}{\includegraphics{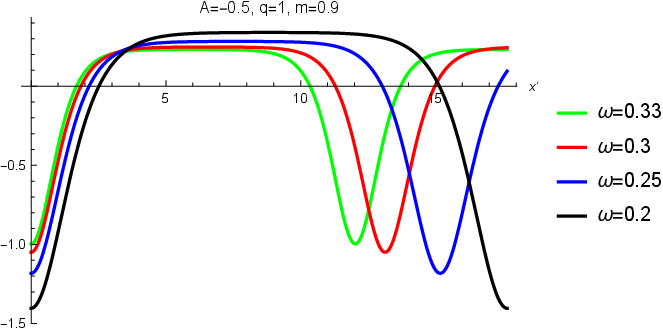}} \\  
\resizebox{0.7\columnwidth}{!}{\includegraphics{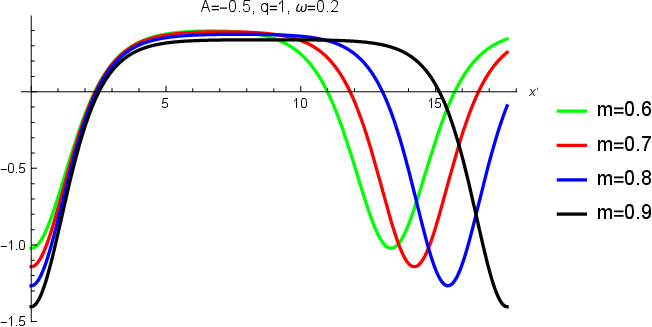}}   
\end{center}
\caption{Top: The profile of the wave (\ref{supEta}) for $A=-\frac{1}{2}, q=1, \omega =0.2, m=0.9$ and its components.  Middle: Profiles of the wave (\ref{supEta}) for $A=-\frac{1}{2}, q=1, m=0.9$ and 
four values of $\omega=0.33, 0.3, 0.25, 0.2$. Bottom: The same for $A=\frac{1}{2}, \omega=0.2$ and four values of $m=0.6,0.7,0.8,0.9$. Note that the black curve represents the same wave.}  \label{f3sup}
\end{figure}

There exist another branch of solutions for $A<0, ~\omega<\frac{q}{\sqrt{3}}$. Their profiles look similar to those presented above, but the velocity of wave propagation is substantially smaller. In Fig.~\ref{f3sup}, we display several examples of such waves with amplitudes close to 1.

\begin{figure}[t]
\begin{center}
\resizebox{0.49\columnwidth}{!}{\includegraphics{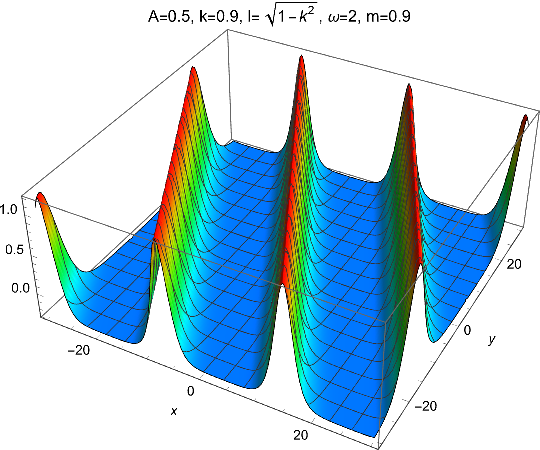}} 
\resizebox{0.49\columnwidth}{!}{\includegraphics{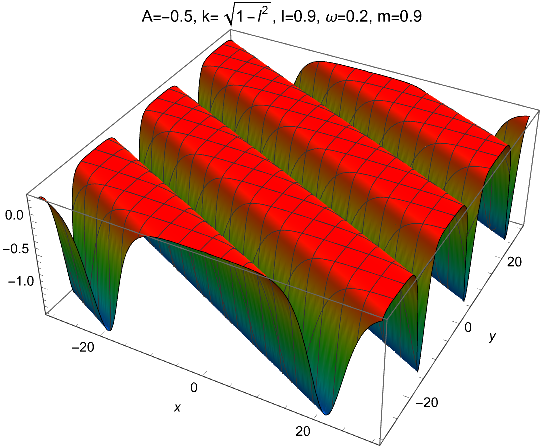}} 
\end{center}
\caption{Left: The 3D-profile of the wave (\ref{supEta}) for $A=\frac{1}{2}, k=0.9, l=\sqrt{1-k^{2}}, \omega =2, m=0.9$. The 1D-profile of the same wave is presented by the black curve in Fig.~\ref{f2sup} along the direction of the wave propagation.
Right: The 3D-profile of the wave (\ref{supEta}) for $A=-\frac{1}{2}, k=\sqrt{1-l^{2}}, l=0.9, \omega =0.2, m=0.9$. The 1D-profile of the same wave is presented by the black curve in Fig.~\ref{f3sup} (middle and bottom) along the direction of the wave propagation.} \label{f3-G-A}
\end{figure}

Finally, we show examples of waves (\ref{supEta}) in full 3D drawings in Figure \ref{f3-G-A}. In both plots $q=1, m=0.9$ were chosen. In the left diagram $A=\frac{1}{2}, k=0.9$ and $\omega=2$, whereas in the right one $A=-\frac{1}{2}, l=0.9$ and $\omega=0.2$. Translation symmetry along the direction perpendicular to the direction of wave propagation is seen.

We see that the superposition solutions to the equation (\ref{rf1b}) admit, in addition to solitary and cnoidal waves presented in previous sections, \emph{table top periodic waves} (we propose this name in analogy to \emph{table top solitons}).

\section{Conclusions} \label{concl}
In this paper, we have obtained approximate solutions to the first-order (2+1)-dimensional  Boussinesq's equations arising from the Euler equations (with appropriate boundary conditions) for an ideal fluid. This time, we assumed equal scaling of the $x,y$ coordinates (meaning equal wavelength in each direction). Unlike our previous work \cite{KRcnsns23,RKappl24,RK2023NoDy,RK25WaveM}, in which the scaling of the $x$ and $y$ coordinates was different (so that the small $\gamma$ parameter was significantly smaller than the $\alpha$ or $\beta$ parameter), in the present work it was not possible to make the Boussinesq equations compatible like in one-dimensional theory. Nevertheless, we obtained traveling wave solutions, analogous to solutions of equations with one spatial dimension. These include soliton solutions, periodic cnoidal solutions and periodic superposition solutions. In this way, we have closed the generalisations of the KdV-type equations to (2+1)-dimensions that follow from the Euler equations for an ideal fluid as first- and second-order approximations.

The results of our previous papers \cite{KRcnsns23,RKappl24,RK2023NoDy,RK25WaveM} and the present study allow the following conclusion. If we restrict ourselves to small amplitude waves, for which the nonlinear equations obtained from the Euler equations for an ideal fluid should be a good approximation, then the solutions of the (2+1)-dimensional equations are analogous to the solutions of the (1+1)-dimensional equations.

\end{document}